\documentclass{ws-ijmpd}
\usepackage{aas}
\usepackage[super,compress]{cite}
\usepackage[breaklinks]{hyperref}
\hypersetup{colorlinks,urlcolor=black,citecolor=black,linkcolor=black,filecolor=black}
\usepackage{breakurl}
\usepackage{url}
\usepackage{subfigure}
\begin{document}

\markboth{Authors' Names}
{Instructions for Typing Manuscripts (Paper's Title)}

\catchline{}{}{}{}{}
%
%%%%%%%%%%%%%%%%%%%%%%%%%%%%%%%%%%%%%%%%%%%%%%%%%%%%%%%%%%%%%%%%%%%%

\title{COSMIC RAY DRIVEN GALACTIC WINDS}

\author{SARAH RECCHIA}
\address{Université de Paris,
CNRS, Astroparticule et Cosmologie, F-75013 Paris, France,\\
recchia@apc.in2p3.fr}

\maketitle

\begin{history}
\received{Day Month Year}
\revised{Day Month Year}
\end{history}

\begin{abstract}

Galactic winds constitute a primary feedback process in the ecology and evolution of galaxies. They are ubiquitously observed and exhibit a rich phenomenology, whose origin is actively investigated both theoretically and observationally.   Cosmic rays have been widely recognized as a possible driving agent of galactic winds, especially in Milky-Way like galaxies. The formation of cosmic ray-driven winds is intimately connected with the microphysics of the cosmic ray transport in galaxies, making it an intrinsically non-linear and  multiscale phenomenon. In this complex interplay the cosmic ray distribution affects the wind launching and, in turns, is shaped by the presence of  winds.\\
In this review we summarize  the present knowledge of the physics  of cosmic rays  involved in the wind formation and of the wind hydrodynamics. We also  discuss the theoretical difficulties connected with the study of cosmic ray-driven winds and  possible future improvements and directions.

\end{abstract}


\ccode{PACS numbers:}

%\tableofcontents

\section{Introduction}

Galactic winds (GWs) are powerful outflows that originate in the disk of galaxies and may extend for hundreds of parsecs.  The ejected material is mainly composed of very hot dilute gas, often enriched of metals, dust and portions of cooler gas,  whose velocity can reach hundreds or even thousands km/s, and, in some cases, the gas may even escape the galaxy and be injected in the intergalactic medium (IGM). The induced  mass loss can be comparable to the mass formed in stars.   

Since their first detection in the M82 starburst 
galaxy\cite{M82-wind-I, M82-wind-II}, 
GWs are ubiquitously observed both locally and at high redshift, especially in galaxies that exhibit a high star formation rate (SFR) or an active galactic nucleus (AGN)\cite{Veilleux-review, Zhang-review, Heckman-review, Rupke-review}.
Also the Milky Way (MW) might host GWs, as suggested by several observations\cite{Miller-2013, Miller-2015}.
GWs represent a most important feedback process which deeply affects the evolution and the structure  of galaxies, star formation, and the chemical and physical properties of the interstellar medium (ISM) and of the IGM.  

The  formation of GWs is not yet fully understood and it is  actively investigated in astrophysics. 
GWs may be powered by the energy and momentum injected by stellar winds and supernova  explosions\cite{Chevalier-Clegg} or by the energy released during accretion onto supermassive black holes in AGN\cite{King-Pound-AGN}.  
They may also be produced by the radiation pressure induced by the scattering/absorption of starlight radiation on dust grains and gas\cite{Scoville-AGN, Murray-Momentum-winds}.
Thermal and radiation pressure gradients are effective in driving winds in  very active galaxies, while in more quiescent galaxies, such as the MW, they are probably not large enough, except maybe for the Galactic Center region. 

On the other hand,  cosmic rays (CRs) escaping from galaxies can effectively push  on the ISM, and eventually launch winds. In fact, CRs are known to be produced in astrophysical sources,  likely supernova remnants (SNRs), located in the galactic disk and to undergo diffusive propagation in the ISM as a result of the scattering off plasma waves. Such scattering provides a strong coupling between CRs and the background plasma, which can allow CRs to produce GWs. 
In the MW CRs are observed  to be energetically  in equipartion with the thermal plasma and with the magnetic field, which  suggests their importance in the dynamics of the ISM. The relativistic CR gas  exhibits a smaller adiabatic index compared to the thermal plasma, so that the CR pressure can  act at relatively large distances from the disk, making possible the wind launching also in situations where the thermal and radiation pressure would fail. 
This aspect has been pointed out for the first time in Ref.~\refcite{Ipavich-1975}, which presented the first study of the CR impact on the launching of GWs in a MW-like galaxy. 
The dynamical role played by CRs in the wind hydrodynamics has been subsequently confirmed by many authors and widely investigated in the stationary  regime \cite{Breitschwerdt-1991-I, Breitschwerdt-1993-II, Everett-2008, Zirakashvili-1996-wind-rot-gal}, in time dependent approaches \cite{Dorfi-2012} and in simulations \cite{Uhlig-2012, Booth-2013, Salem-2014, Girichidis-2016, Ruszkowski-2016, Pakmor-2016-wind-diffusion-simu, Pakmor-2016-wind-cr-transport-simu, Simpson-2016-cr-pressure-winds-simu, Wiener-2017-streaming-diffusion-simu, Pfrommer-2017-simu, Fujita-2018-wind-simu, Farber-2018-wind-simu, Jacob-2018-winds-halo-mass-simu, Thomas-2019-cr-hydro-simu}.

The possibility of CRs to launch winds is intrinsically connected with their transport in the ISM, which in turn is affected by the presence of GWs.
In fact, CRs escaping from the source region, namely the disk, establish a gradient in their distribution function whose consequences are manifold. First of all, such gradient can lead to the excitation of plasma instabilities, most notably of the streaming instability of Alfvén waves\cite{Kulsrud-1971, Skilling-1975-III}, which in turn affects the scattering of CRs, namely their diffusive motion, and their coupling with the background gas.  
Moreover, the coupling with the ISM and the CR gradient between the disk and the outer halo may induce the formation of GWs, which in turn affect the advective transport of CRs. Finally,  the damping of the CR-induced plasma turbulence and the ionization losses of sub-GeV CRs can heat the gas\cite{Walker-2016}, further affecting the wind launching and the  wind structure. 
Thus, the dynamics of CR-driven winds is an intrinsically non linear process and a multiscale phenomenon, in which the microphysics of  the CR transport, on scales of the Larmor radius of relativistic particles in the galactic magnetic field,   determines the CR distribution function and the wind structure,  which in turn affect the CR propagation itself.

The multiscale and nonlinear nature of CR-driven winds makes them particularly difficult to study, especially when CRs are not treated macroscopically, as a relativistic gas, but the details of the CR microphysics is taken into account \cite{Ptuskin-wind-transp, Recchia-winds-I, Recchia-winds-II}.\\

In this review we summarize the relevant properties and the physics involved in the formation CR-driven winds. 
The review is organized as follows: 1) in Sec.~\ref{sec:gal-winds} we review the observations, formation mechanisms and the impact of GWs in different types of galaxies; 2) in Sec.~\ref{sec:CR-nutshell} we briefly introduce the main aspects of CRs astrophysics; 3) in Sec.~\ref{sec:wind-CR-transp} we illustrate the relevant ingredients of the CR transport in a CR-driven wind; 4) in Sec.~\ref{sec:wind-hydro} we review models of GW hydrodynamics in the presence of CRs; 5) in Sec.~\ref{sec:wind-stat} we show in some detail the properties of CR-driven winds in the stationary regime, in particular in  a MW-like galaxy; 6) in Sec.~\ref{sec:concl} we drive some conclusions and discuss  future perspectives. 

%%%%%%%%%%%%%%%%%%%%%%%%%%%%%%%%%%%%%%%%%%%%%%%%%%
%%%%%%%%%%%%%%%%%%%%%%%%%%%%%%%%%%%%%%%%%%%%%%%%%%
%%%%%%%%%%%%%%%%%%%%%%%%%%%%%%%%%%%%%%%%%%%%%%%%%%

\section{Galactic winds}
\label{sec:gal-winds}

Different gas  phases coexist in GWs: dilute hot and warm phases (very hot $\sim\, 10^8$ K, hot $\sim\, 10^6 - 10^7$ K and warm $\sim\, 10^4$ K), neutral atomic matter  ($\sim\, 10^3$ K), cold molecular material, dust ($\sim\, 100$ K) and non-thermal particles. 
Both hot and warm/cool gas can  exhibit  very high velocities (hundreds or thousands km/s), and produce  mass loss rates which can be as large as several solar masses per yr. How these phases are accelerated to such high velocities is being actively investigated in astrophysics and several aspects  remain unclear. Moreover, the multiphase nature of GWs and the fact that a large part of the ouflowing gas is hot and very dilute,  make their observation   challenging.   

In this section we briefly  summarize  the  mechanisms responsible for the wind formation in different types of galaxies and environments, and the observational techniques employed to detect and study GWs. We conclude by stressing the importance and implications of GWs in astrophysics and cosmology. For further details we refer the reader to the excellent reviews found in Refs.~\refcite{Veilleux-review},\refcite{Zhang-review}, 
\refcite{Heckman-review} and \refcite{Rupke-review}.

\subsection{Production mechanisms}

Hot winds  are most likely accelerated in starburst regions or in AGN. The mechanical energy injected in such regions and partially  converted to thermal energy through shocks,  creates a bubble of very  hot gas whose pressure can be much larger than that of the surrounding ISM. The dynamical evolution of such bubbles has been widely studied, starting form the seminal  work of Ref.~\refcite{Chevalier-Clegg}, which showed that, under the assumption that  radiative losses, gravity  and other effects can be neglected,  the flow solution  is  self-similar. 
The bubble may then be able to cross the disk height and eventually break through the galactic halo. 
In general, in such energy-driven flows,  the energy input must be large enough, or radiative losses small enough, that the wind is not hampered  before it can reach the edge of the galactic disk. 

Winds may also be  driven by the mechanical  momentum imparted by starburts and AGN\cite{Scoville-AGN, King-Pound-AGN} or by radiation.
Radiation can  heat the ISM via photoionization. For instance, UV photons can create  regions of ionized hydrogen with temperatures of $\sim 10^4$ K. 
In addition, radiation pressure on dust grains and gas\cite{Scoville-AGN, Murray-Momentum-winds}  can effectively  push the ISM and contribute to the generation of GWs.
Radiation pressure may be important in different environments, from massive stars to luminous AGNs, from starburst  to  rapidly star-forming galaxies\cite{HAM-1990-review, Veilleux-review, Zhang-review, Heckman-review}. 

A very important and still unclear aspect is how the warm and cold wind phases interact with the hot phase and are themselves accelerated to velocities that are comparable to that of the hot gas, and that can be as high as hundreds km/s. A possible scenario is that portions of the warm and cold ISM, or even clouds, are advected in the hot wind and accelerated by its strong ram pressure\cite{Zhang-review, Heckman-review}. Another possibility is that the warm/cool gas in GW  forms due to the thermal instabilities induced in the gas by radiative cooling\cite{Thompson-multiphase-wind, Zhang-review, Heckman-review} 

\subsection{Observations}

Given the multiphase nature of GWs,  multiwavelength observations, of both continuous and line emission and of absorption lines, spanning from radio to X-rays, are necessary  in order to identify and characterize the various flow phases \cite{HAM-1990-review,
Veilleux-review, Rupke-review}.\\
The flow velocity is strongly phase dependent, with the hotter phases showing the highest speeds (up to thousands km/s), down to the $\sim 100$ km/s found in the cold  molecular and atomic gas phases\cite{Heckman-review}. The study of the line profile of emission and absorption lines allows to estimate the fluid velocity, while spatially resolved mapping of the emitting/absorbing material is used to determine the wind morphology\cite{HAM-1990-review, Rupke-review}. 

The hot and very hot  gas phases are mainly tracked through their X-ray emission. In particular,  hard X-rays  are expected 
from the very hot and tenuous gas heated by SNe and stellar winds, and from internal shocks in such hot wind. When the wind collides with  denser ambient material or with clouds, the shock heated/evaporated material is expected to produce also soft X-rays\cite{HAM-1990-review}.

The interaction of a hot wind with a denser and cooler background gas, possibly characterized by a mixture of phases, can produce radiative shocks that compress and heat the ambient medium , thus producing the warm wind phases. 
Such warm phases  have been studied through the  
continuous  emission in the vacuum ultraviolet and through  emission lines, in the optical band (e.g $\rm H\alpha$, NII, OII, OIII, and SII) and in the near-IR band (e.g the $\rm 2.122\,\mu m$ line of $\rm H_2$, and the lines $\rm Pa\alpha$ and $\rm Br\gamma$)\cite{Heckman-review,
Veilleux-review}. The presence of shocks is often revealed by the presence of double-peaked emission with typical values of emission-line ratios like SII/$\rm H_{\alpha}$, NII/$\rm H_{\alpha}$ and OII/$\rm H_{\alpha}$\cite{HAM-1990-review, Veilleux-review}.

Molecular gas is mapped through mm-wave (warm) and mid/near IR (cold) spectroscopy. Atomic hydrogen is tracked by the 21 cm line, but also  OI and CII lines in the far IR are used to identify  atomic gas.  Finally, dust is studied both in the IR and in the far-UV band\cite{Heckman-review}.

Radio surveys reveal the presence of synchrotron emitting relativistic electrons   in magnetized winds\cite{Heckman-review}.

Complementary to  emission techniques,  absorption-line techniques are also massively employed, especially for the study of face-on systems and of distant galaxies, where the diffuse emission from the wind is hard to detect. 
The most used lines are in the UV band: Silicon (Si II $\lambda 1260$, $\lambda 1526$, OI + SiII $\lambda 1303$), Carbon (CII $\lambda 1334$), Iron (FeII $\lambda 1608 $) and Aluminum (Al II $\lambda 1670$), Oxygen (OVI $\lambda\lambda 1032,\, 1038 $), and in the optical band: Na I D ($\lambda\lambda 5890,\, 5896 $), KI  ($\lambda\lambda 7665,\, 7699 $)\cite{Veilleux-review}.\\

As for the Milky Way, the existence of GWs has not  been firmly  established. However,  the recent observation in the X-ray band of Oxygen OVII and OVIII absorption lines in the spectra of distant quasars\cite{Nicastro-2012-gas-halo, Miller-2013, Miller-2015}, implies the existence around the Galaxy of a very extended ($\sim$ 100-200 kpc) halo constituted of hot ($\sim~10^6$ K) dilute  gas, with metallicity larger than 0.3, and likely connected with GWs. Moreover, also the so called Fermi Bubbles, giant structures  detected in $\gamma-$rays that extends for several kpc above and below  the Galactic Center,  might be the result of outflows\cite{Lacki-Fermi-Bubbles, Cheng--Fermi-Bubbles, Zubovas-Fermi-Bubbles}.

\subsection{Impact of galactic winds}

GWs constitute a most relevant ingredient in the ecology of galaxies, and a primary feedback mechanism.\\
In fact, GWs heat the ISM and drain hot gas and metals from the disk of the host galaxy  toward  galactic halos or even in the intergalactic space.
This limits the amount of gas available to form stars. Moreover, the injected material affects the surrounding  space by changing the temperature, the degree of ionization and the  chemical composition, which especially  influence  the cooling rate\cite{Dalgarno-1972, Miller-2015}, and so  the time required for the ejected gas  to fall back to the disk. 
All this establishes a gas recycling  and a regulation of the SFR\cite{Veilleux-review, Zhang-review, Heckman-review}. Models of galactic evolution that do not include GWs tend  to overestimate  the amount of baryons and the SFR\cite{Crain-2007}. 
GWs are necessary in order to explain the luminosity function of galaxies and the mass/metallicity relation (see Ref.~\refcite{Zhang-review} and references therein).\\
In the cases of dense outflows, star formation may even take place within the outflow, with prominent consequences  on the spheroidal component of galaxies and on the chemistry of the circumgalactic and intergalactic medium\cite{Maiolino-2017-star-form-wind}.\\
The gas expelled by GWs in the IGM may represent a sizable fraction of the baryonic matter of the Universe\cite{Kalberla-2008A&A...487..951K, Miller-2015} and has also been invoked as a possible important contribution to the   solution  of the missing baryons problem in the local Universe\cite{Michael-2010}.\\
Finally GWs may also transport magnetic turbulence from the disk to the halo  and may contribute to shaping the large scale magnetic field in galaxies, especially the out of disk components\cite{Hanasz-2017}.

%%%%%%%%%%%%%%%%%%%%%%%%%%%%%%%%%%%%%%%%%%%%%%%%%%%%%%%%
%%%%%%%%%%%%%%%%%%%%%%%%%%%%%%%%%%%%%%%%%%%%%%%%%%%%%%%%
%%%%%%%%%%%%%%%%%%%%%%%%%%%%%%%%%%%%%%%%%%%%%%%%%%%%%%%%
%%%%%%%%%%%%%%%%%%%%%%%%%%%%%%%%%%%%%%%%%%%%%%%%%%%%%%%%
%%%%%%%%%%%%%%%%%%%%%%%%%%%%%%%%%%%%%%%%%%%%%%%%%%%%%%%%

\section{Cosmic rays in a nutshell}
\label{sec:CR-nutshell}

\begin{figure}[h!]
	\centering
	\includegraphics[width=\columnwidth]{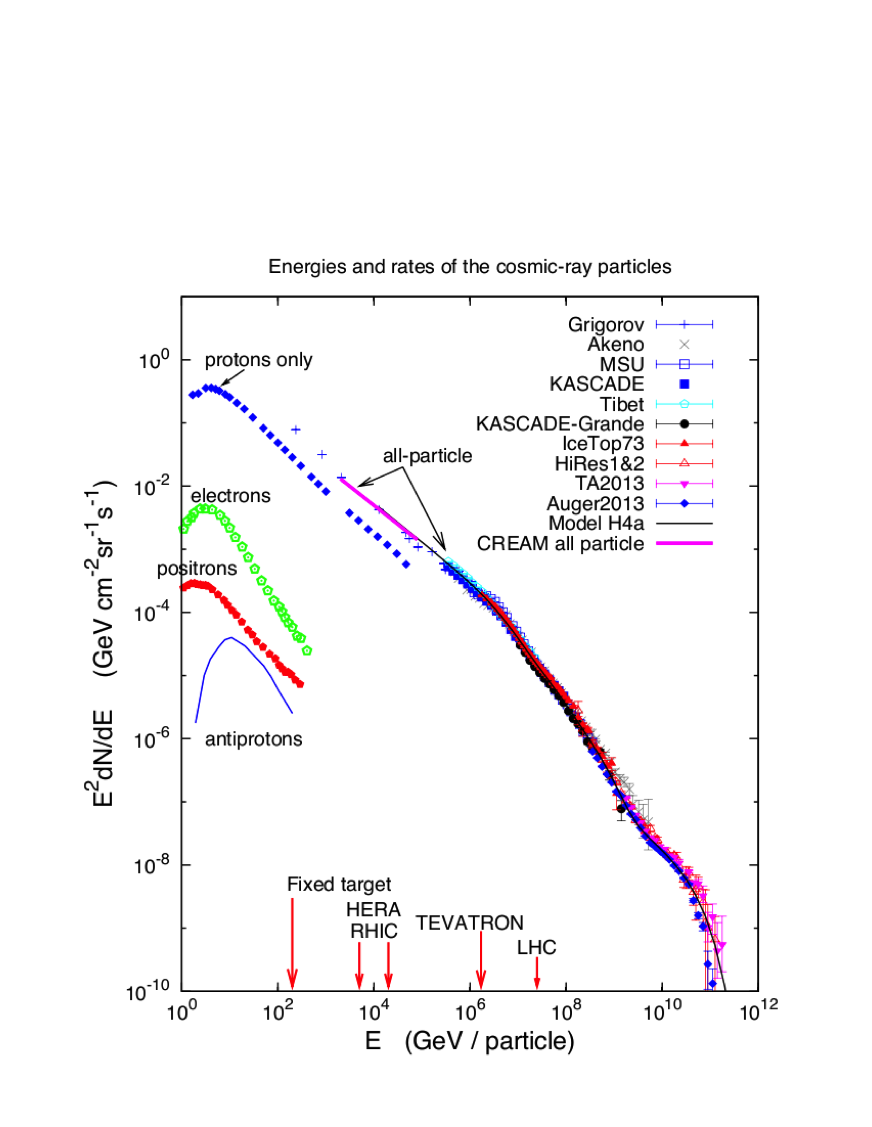}
    \caption{CR spectrum at Earth as detected by different experiments\cite{Hillas-2006} (courtesy Tom Gaisser).}
	\label{fig:all-spectrum}
\end{figure}

CRs are very high energy charged particles that originate in astrophysical sources and reach our planet by propagating and interacting with the ISM. Locally, CRs are revealed with balloon, satellite and air-shower detectors. The observed CR flux is shown in Fig.~\ref{fig:all-spectrum}. Remarkably, it extends as a nearly perfect power-law in energy, $\rm \sim E^{-\gamma} $, from GeV energies  up to $\sim 10^{20}$ eV, with a  corresponding energy density  of $\rm \sim 1~eV/cm^{-3}$, a value comparable with the thermal and magnetic pressure observed in the Galactic disk. 
Protons largely dominate the observed flux, followed by Helium ($\sim 10\%$) and heavier nuclei ($\sim 1\%$). The leptonic flux is about 100 times smaller than the  hadronic flux, and electrons reaching $\sim 20$ TeV have been measured so far\cite{HESS-2017-HEE, Recchia-2019-TeV-elec}. 
Notable spectral breaks occur at:  $\rm \sim  230\, GV$, more pronounced in the proton and Helium spectra, but present also for heavier nuclei,  where the slope of the proton spectrum changes from $\gamma = 2.89 \pm 0.015 $ to $\gamma = 2.67 \pm 0.03 $\cite{pamhard, amshard}; the \textit{knee}, at $\approx 3\times 10^{15}$ eV, where the spectral slope changes from $\gamma \approx  2.7$ to $\gamma \approx  3.1$ and the chemical
composition seems to become heavier; the \textit{ankle}, at $\approx 3\times 10^{18}$ eV, where the spectral slope flattens to  $\gamma \approx  2.7$\cite{Hoorandel-2006-review-knee}.
%%%%%%

CRs are also  tracked through the radiation produced by their interaction with the ISM, most notably through  $\gamma$-rays  produced by the decay of $\pi^0$ mesons generated by the interaction of CR protons with the background gas, and by bremsstrahlung and inverse Compton due to CR leptons, and by synchrotron radiation  produced by the interaction of CR leptons with the background magnetic field.
%%%%%%%%
\begin{figure}[h!]
	\centering
	\includegraphics[width=\columnwidth]{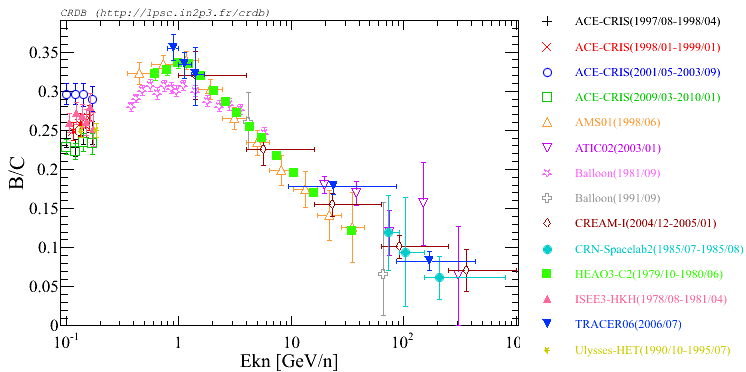}
    \caption{Boron over Carbon ratio as a function of the kinetic  energy per nucleon\cite{Maurin-2014-database}.}
	\label{fig:BC}
\end{figure}

Together with the flux, the  chemical composition and the level of anisotropy of the observed CRs,  represent precious pieces of information. In particular, the relative abundance of secondary spallation nuclei relative to their
primary nuclei (e.g the Boron/Carbon ratio, see Fig.~\ref{fig:BC}, or the abundance of  unstable isotopes relative to stable isotopes (e.g $\rm {}^{10}Be/{}^{9}Be$), represent the most reliable measurement of the amount of matter  traversed by CRs during their travel through the ISM, and so of the residence time of CRs in the Galaxy. The estimated residence time decreases with energy and   exceeds by orders of magnitudes the ballistic propagation time. This strongly supports the idea that CRs propagate diffusively in the magnetized ISM, as additionally confirmed by  the small observed anisotropy ($\sim~ 0.001$ at TeV energies)\cite{Blasi-2013-review, Gabici-2019-review-con-altri}.\\

Nowadays, the most accredited  picture of the origin of Galactic CRs, is the so-called \textit{supernova remnant paradigm}\cite{Berezinskii-1990, Blasi-2013-review, Gabici-2019-review-con-altri}. In this scenario, the bulk of CRs is
thought to be accelerated in the Galactic disk at the shocks of supernova remnants (SNRs) through the mechanism of diffusive shock acceleration (DSA).
On the energetic ground, this requires that 3-10\% of the mechanical energy released by
SNe  is channeled into CRs\cite{Baade1934-SN, Ginzburg-1961PThPS..20....1G}. 
DSA  is  rigidity dependent and produces  CRs with approximately universal power law energy spectra $\rm \approx E^{-2.0}$. Protons are thought to be accelerated up to energies  $\gtrsim 3\times 10^{15}$ eV, while nuclei with charge Z would reach an energy Z times larger, namely a factor of 26 for Iron nuclei. In this picture the \textit{knee} would naturally result from the superposition of the energy cut-offs of the different nuclear species.
Radio, X-ray and $\gamma-$ray observations confirmed SNR shocks as acceleration sites of CRs\cite{Vink-2012-SNRs-acc-obs}, however it is still unclear whether they can accelerate CRs up to PeV energies (see Refs.~\refcite{Blasi-2013-review}, \refcite{Gabici-2019-review-con-altri} and references therein).

After leaving their sources, CRs are thought undergo diffusive propagation in the turbulent Galactic magnetic filed. Magnetic inhomogeneities can   scatter CRs and produce   their spatial diffusion. The scattering occurs on scales of the Larmor radius $r_L$  of the particles and is likely  of resonant nature, namely  CR with a given larmor radius are scattered by plasma waves  whose wave number $k$ satisfies the relation $k \sim 1/r_L$\cite{Jokipii-1966, Kulsrud-2005}.
The decrease of the Boron/Carbon ratio with energy can be accounted for by a diffusion coefficient that increases with energy.  The spectral hardening at $\sim 230 $ GV may be due  to a change in the type of  turbulence relevant for CR scattering  at $\sim 200-300$ GV  \cite{Aloisio-2013, Aloisio-2015}, but other possible  interpretations have been proposed (see Ref.~\cite{Recchia-2018-pamela}).

The origin of CRs beyond the \textit{ankle} is probably extragalactic, since the Larmor radius of such high energy particles in the typical Galactic magnetic field is comparable to the size of the Galaxy. On the other hand, the energy of transition between Galactic and extragalactic CRs is still largely debated\cite{Blasi-2013-review, Gabici-2019-review-con-altri}.

\subsection{Cosmic ray propagation: building blocks}
\label{subsec:CR-prop-basics}

The  propagation of CRs in the Galaxy is an intricate, non-linear,  phenomenon which depends on several processes that are often not understood. The details of  CR propagation are intimately connected with the formation of CR-driven winds. 

CRs are thought  to undergo a diffusive motion during their escape from the Galaxy  due to the scattering on magnetic waves, as suggested by the measured abundances of secondary nuclei and unstable isotopes and by the observed small level of anisotropy. In addition, CRs may be advected out of the Galaxy due to the plasma waves themselves and to large scale outflows, such as GWs. Finally, other 
processes, such as ionization and Coulomb losses\cite{Walker-2016}, bremsstrahlung, synchrotron and inverse Compton, turbulent reaccelaration, nuclear fragmentation and radioactive decay, proton-proton interactions, may affect the transport, depending on the CR energy and species\cite{Berezinskii-1990}.

All these processes depend on the specific conditions of the ISM, such as the density, temperature and degree of ionization of the background medium, the magnitude and the level of turbulence of the magnetic field, the background photon field, and, in the case of GWs, also on the Galactic gravitational potential. 
Such conditions can vary from point to point in the Galaxy and CRs can play an active role in determining the propagation parameters\cite{Recchia-winds-II}.\\ 

The last aspect is particularly relevant in the case of CR-driven GWs. 
In fact,  the CR density gradient that exists between the Galactic disk, where the sources of the bulk of CRs are probably located, and the halo, together with  the fact that  CRs are observed to be roughly in equipartition with the thermal plasma and with the background magnetic field, suggest that CRs may play a relevant dynamical role.\\
It is known that spatial gradients in the CR density can lead to the excitation of Alfv\'{e}n waves, at the CR Larmor radius scale, through the so called \textit{streaming instability}\cite{Kulsrud-1969, Skilling-1975-III}, which propagates along the magnetic field lines in the direction of the decreasing CR density. 
The scattering of CRs on such waves affects their diffusive propagation and contribute to  determine the  diffusion coefficient, together with other sources of magnetic turbulence. 
The  level of plasma waves is limited due to different damping processes, such as the turbulent cascading\cite{Zhou-cascade, Miller-1995-cascade},    the non linear Landau damping\cite{Kulsrud-1978-nlld, Wiener-2013-WIM-ion-neutral}, the turbulent damping\cite{Farmer-2004, Yan-Lazarian-2004-compressible} and the ion-neutral damping\cite{Zweibel-1982-ion-neutral}, depending on the conditions of the background medium. 
A larger CR gradient leads to a  higher level of turbulence, thus to a stronger CR scattering and to a smaller diffusion coefficient.
Moreover, in the case of self-generated Alfv\'{e}n waves, that 
move in the direction opposite to the CR gradient,  CRs experience also convection with such waves at roughly the Alfv\'{e}n speed. 
It has been shown\cite{Blasi-2012PhRvL.109f1101B, Aloisio-2013} that streaming instability of Alfv\'{e}n waves may dominate the CR propagation up to $\sim 200$ GeV, while at higher energies other sources of turbulence are probably at work\cite{Aloisio-2015}.
The nature of such additional types of turbulence  is still debated. In fact, for a long time Alfv\'{e}nic turbulence injected on large scale ($10-100$ pc) by astrophysical sources, such as SNe, and cascading to the resonant scales, was thought to be responsible for the CR scattering. However, subsequent studies showed that   Alfv\'{e}nic turbulence cascading to small scales becomes anisotropic and very inefficient at scattering CRs\cite{GS-1994-I, GS-1995-II, Chandran-2000}.
This problem does not affect self-generated Alfv\'{e}n waves, which are already produced at the resonant scale.   Recently it has been suggested that compressive MHD waves could provide a relevant contribution to CR scattering\cite{Yan-Lazarian-2004-compressible, Lazarian-2006-compressive-MHD, Evoli-Yan-2014}.

In most cases relevant for galactic propagation, the level of magnetic turbulence on scales resonant with the Larmor radius of CRs is $\rm \langle \delta B \rangle /B \ll 1$ and diffusion can be treated in the framework of quasi-linear theory\cite{Berezinskii-1990, Kulsrud-2005}.
This implies that CRs diffusion is  anisotropic, with particles diffusing preferentially
along rather than across the magnetic field lines, 
which translates in a parallel and perpendicular diffusion coefficient $D_{\parallel} \gg D_{\perp}$.
In this case, the CR transport perpendicular to the mean background field may be provided by the wandering of field lines\cite{Jokipii-1966, Casse-2002-chaotoc-B, Nava-2013-anisotropic}.

The coupling established by CRs and the ISM by scattering and the presence of a  CR gradient between the disk and the halo, imply that CRs  exert a force on the background plasma  directed toward the Galactic halo. If this force is large enough as to win against the gravitational force, a GW may be launched. The hydrodynamics of such a wind is determined by CRs. On the other hand, the advective transport of CRs is influenced by the presence of a wind. 

Since CR protons are the most abundant species, all these phenomena, that depend on the CR density gradient, are mostly determined by protons. Moreover, the wind launching depends on the CR pressure, which is dominated by 1-10 GeV protons.

\subsubsection{Basic model of  proton transport}

The propagation  phenomenology described above is very rich and complicated. 
However it is possible to understand some basic aspects of CR propagation relevant in the case  CR-driven winds,  with the help of a  simplified  stationary model of proton transport. 
In such basic model the CR propagation is assumed to be 1D and to occur only perpendicular to  the Galactic disk and solely along a constant background magnetic field. The diffusion coefficient is assumed to be spatially homogeneous and is treated as a fitting parameter. The specific nature of the turbulence responsible for the CR scattering is ignored. The injection of CRs occurs in  the Galactic disk, assumed to be infinitely thin, while  also the advection velocity (assumed to be constant) and the size of the propagation region are preassigned. Other possible effects, such as ionization and Coulomb losses, are neglected.\\
In this scenario the stationary CR advection-diffusion equation reads\cite{Blasi-2013-review}
\begin{equation}\label{eq:slab-eq}
-\frac{\partial}{\partial z} \left[D\frac{\partial f}{\partial z} \right] + u\frac{\partial f}{\partial z} -
\frac{d\, u}{d\, z}\frac{p}{3}\frac{\partial f}{\partial p} = Q(z,p),
\end{equation}
where $f(z,p)$ is the CR distribution function, which is function of the distance from the Galactic disk $z$ and of the particle momentum $p$. 
Under the following assumptions this equation is easily solved: the diffusion coefficient $D(p) \propto p^{\delta} $ is a power-law in  momentum, the injection occurs in the disk  $Q(z,p) = Q_0(p)\delta(z)$ with a power law  injection spectrum $Q_0(p)\propto p^{-\gamma}$ (as expected for DSA\cite{Blasi-2013-review}), the advection velocity  $u$ is constant  and directed away from the disk ($\frac{d\,u}{d\,z} = 2u\delta(z)$), and CRs freely escape at $\rm \vert z \vert = H$, namely $f(\pm H, p) =0$. The free escape boundary H, namely the size of the propagation region, if fixed, and it must be not infinity in order to guarantee the existence of a stationary solution. 

A standard way to solve Eq.  \ref{eq:slab-eq} is to integrate it around the disk, between $z=0^{-}$ and $z=0^{+}$, and between $z=0^{+}$ and $z$, with the free-escape boundary condition.
The solution reads 
\begin{align}\label{eq:slab-sol}
&  f(z,p) = f_0(p) \frac{1-e^{-\zeta\left(1-\frac{\vert z\vert}{H}\right)}}{1-e^{-\zeta}}\\
& f_0(p) = \frac{3}{2\,u}\int_p^{\infty}\frac{dp^{\prime}}{p^{\prime}}Q_0(p)\exp\left[\int_p^{p^{\prime}}
\frac{d\,p^{\prime\prime}}{p^{\prime\prime}} \frac{3}{\lambda(p^{\prime\prime})}\right],
\end{align}
where $\zeta(p) = \frac{u\,H}{D(p)}$ and $\lambda(p) = 1 - e^{\zeta(p)}$.\\
For particle momenta for which advection dominates, namely $\frac{u\,H}{D(p)} \gg 1$,  the CR spectrum becomes 
\begin{equation}\label{eq:adv-dominate}
    f_0^{adv}(p) = \frac{Q_0(p)}{2\, u} \sim p^{-\gamma}.
\end{equation}
Thus, in this regime the equilibrium  spectrum has the same slope of the injection spectrum.\\
When diffusion dominates, namely $\frac{u\,H}{D(p)} \ll 1$, the CR spectrum becomes 
\begin{equation}\label{eq:diff-dominate}
   f_0^{diff}(p) = \frac{Q_0(p)\, H}{2\,D(p)} \sim p^{-\gamma -\delta}
\end{equation}
and the equilibrium  spectrum is steeper than the injection spectrum due to the energy dependent diffusive propagation. 
Since, as inferred, for instance, from the  the Boron/Carbon ratio, the diffusion coefficient is an increasing function of momentum, advection(diffusion) dominates for low(high) enough particle momenta.
Typical values used to fit CR data are: slope of the  injection spectrum $\gamma \sim 4.1-4.4$, slope of the diffusion coefficient $\delta \sim 0.3-0.6 $ and normalization $\sim 3\times 10^{28}$ cm$^2$/s at $\sim 1$ GeV, halo size of $H\sim 4$ kpc and  advection velocity of $\sim 20$ km/s\cite{Blasi-2013-review, Aloisio-2015}.\\

It is interesting to notice that in a GW scenario the size of the diffusive region emerges naturally, without the need of imposing a free-escape boundary. 
This is due to the fact that, as we will see in the remaining of this review, in a stationary GW the wind velocity increases with the distance from the disk and eventually tends to a constant value. This means that, for a  particle of momentum $p$, advection starts to dominates over diffusion at a distance from the disk, $s^{*}(p)$, given by
\begin{equation}\label{eq:s-star}
\frac{s^{*2}(p)}{D(s^*,p)} = \frac{s^*(p)}{U(s^*)}.
\end{equation}
Since $D$ generally increases with momentum, the size of the diffusion region also increases with momentum. Above $s^{*}(p^{*})$ particles of momentum $p^{*}$ are basically advected out of the Galaxy.
All this may have relevant  consequences on the CR spectrum, anisotropy and production of secondaries\cite{Ptuskin-wind-transp, Recchia-winds-I}.

%%%%%%%%%%%%%%%%%%%%%%%%%%%%%%%%%%%%%%%%%%%%%%%%%%%%%%%%%%%%%%%%%%%%%%%%%%%%%%%%%%%%%%%%%%%%%%%%%%%%
%%%%%%%%%%%%%%%%%%%%%%%%%%%%%%%%%%%%%%%%%%%%%%%%%%%%%%%%%%%%%%%%%%%%%%%%%%%%%%%%%%%%%%%%%%%%%%%%%%%%
%%%%%%%%%%%%%%%%%%%%%%%%%%%%%%%%%%%%%%%%%%%%%%%%%%%%%%%%%%%%%%%%%%%%%%%%%%%%%%%%%%%%%%%%%%%%%%%%%%%%

\section{Cosmic ray driven winds: cosmic ray transport}
\label{sec:wind-CR-transp}

The inclusion of the CR transport in models of GWs is made extremely difficult by its  non linear nature and by the fact that the  scales relevant for the CR propagation (Larmor radius of the particles) are much smaller than the those relevant for the wind ($\gtrsim$ kpc). 
For this reason, in most wind models and simulations   CRs are treated as a fluid. The first attempt to account for the effects of a stationary CR driven wind on the CR transport was by  Ref.~\refcite{Ptuskin-wind-transp},  where the  authors used a simplified model of the wind structure (as inferred from hydrodynamic calculations) in order  to deduce some general implications
for the observed CR spectrum. More recently Ref.~\refcite{Recchia-winds-I}. presented a semi-analytic, self-consistent calculation, of the wind formation together with the  propagation of CR protons in the wind. The work is restricted to  the stationary regime, and includes the self-generation of Alfv\'{e}n waves by  CR streaming instability, the damping of such waves and the resulting gas heating, the CR scattering parallel to the field lines on the self-generated waves, and the CR advection with such waves and with the wind. The CR pressure contributes to the wind formation and is dominated by CR protons, which are the  most abundant CR species. The authors also performed a parametric study of the wind launching and CR distribution in different galactic environments\cite{Recchia-winds-II}. \\
Based on such studies, here we briefly summarize the main features of the CR transport in a CR-driven wind that have been  explored in the literature, and how they are linked to the wind hydrodynamics, which is described in the next section. Possible  future improvements of the present models are illustrated in the conclusions.\\

In the assumption that the CR transport is dominated by the scattering on  self-generated Alfv\'{e}n waves and in the presence of a GW, the CR diffusion-advection equation reads\cite{Skilling-1975-III}
\begin{equation}
\frac{\partial f}{\partial t} +(\vec{u}+\vec{v}_A)\cdot\vec{\nabla}f= \vec{\nabla}\cdot\left[D\vec{\nabla}f\right]
	+ \left[ \vec{\nabla} \cdot (\vec{u}+\vec{v}_A) \right] \frac{p}{3} \frac{\partial f}{\partial p} + Q_c
\end{equation}
where $f(p, \vec{r})$ is the CR distribution function and $D(p, \vec{r})$ the CR diffusion coefficient, while $\Vec{u}(\vec{r})$ and $\Vec{v_A}(\vec{r})$ are the background plasma velocity (the wind velocity in this context) and the Alfv\'{e}n velocity. 
In the case of self-generation, Alfv\'{e}n waves move preferentially along the magnetic field lines and in the direction of the decreasing CR density, so that CRs will be advected at a speed which is roughly $\vec{u} + \vec{v}_A$.
In the case of an extrinsic turbulence, waves would move along the field lines in both directions and CRs would advect at $\vec{u}$\cite{Skilling-1975-III}.\\
The CR diffusion coefficient is given by the Alfv\'{e}n wave energy density $W(k, \vec{r})$, determined by the CR streaming instability and by the possible sources of wave damping. In the context of galactic propagation the level of turbulence is weak at the scale resonant with CRs and the diffusion coefficient can be expressed using quasi-linear theory\cite{Berezinskii-1990, Kulsrud-2005} :
\begin{equation}\label{eq:D-QLT}
D(\vec{r},p)=\left.\frac{1}{3}\frac{v(p)r_L(\vec{r},p)}{k_{\rm res}W(k_{\rm res}, \vec{r})}\right|_{k_{\rm res}=1/r_L},
\end{equation}
where $v$ and $r_L$ are the particle velocity and Larmor radius, and the wave energy density is calculated at the resonant wavenumber $k_{\rm res}=1/r_L$.\\
The Alfv\'{e}n wave transport equation,  taking into account the velocity of the background medium,  wave generation by CR streaming instability and the relevant damping processes, is given by\cite{Jones-1993ApJ...413..619J}
\begin{equation}
\frac{\partial W}{\partial t} + \vec{\nabla}\cdot\left[\left(\frac{3}{2}\vec{u}+\vec{v}_A\right)W\right] =\frac{1}{2}\vec{u}\cdot \vec{\nabla}W -[\Gamma_{\rm CR}+ \Gamma_{\rm damp}]W
\end{equation}
where $\Gamma_{\rm CR}$ is the wave growth rate due to CR streaming instability (the $\delta-$function encloses the resonance condition):
\begin{equation}
\Gamma_{\rm CR}=\frac{16\pi^2}{3}\frac{\vec{v}_A}{kWB^2}\cdot\int_0^\infty dp p^4 v(p)\delta\left(p- \frac{qB}{kc}\right)\vec{\nabla}f,
\end{equation}
and $\Gamma_{\rm damp}$ is the wave damping rate.\\
The damping of waves depends on the conditions of the background plasma and on the wave density itself. In a hot ionized  medium the most relevant processes are the turbulent cascade, the non linear Landau damping and the turbulent damping (see Sec.~\ref{subsec:CR-prop-basics}).\\
The turbulent cascade of waves to smaller scales is a nonlinear process that can be described as a diffusion in wavenumber space, with a diffusion coefficient given by\cite{Zhou-cascade, Miller-1995-cascade}
\begin{equation}
    D_{kk} = 5.2\times 10^{-2} v_A k^{7/2}W^{1/2}.
\end{equation}
At small enough scales waves will eventually dissipate and heat the gas. The typical  rate for this process is $\Gamma_{\rm casc} \sim D_{kk}/k^2$.\\
Non linear Landau  damping is due the interaction between the beat of two Alfv\'{e}n waves and the thermal ions, of temperature $T$,  in the background plasma. The damping rate is given by\cite{Kulsrud-1978-nlld,  Wiener-2013-WIM-ion-neutral}
\begin{equation}
    \Gamma_{\rm NLL} = \frac{1}{2}\sqrt{\frac{\pi}{2}\frac{k_{\rm Bol}\,T}{m_p}}\frac{k\,W}{r_{L}}.  
\end{equation}
Turbulent damping is due to the interaction of the self-generated Alfv\'{e}n waves with the background turbulence. Such turbulence can be injected by astrophysical sources at scales, $\rm L_{inj}$, much larger than that of the Larmor radius of CR particles, and with a turbulent velocity $\rm v_T$. For waves resonant with CRs with a given Larmor radius, the damping rate is given by\cite{Farmer-2004, Yan-Lazarian-2004-compressible}
\begin{equation}
    \Gamma_{\rm turb} = \left(\frac{v_T^3/L_{inj}}{r_L\, v_A} \right)^{1/2}.
\end{equation}
%%%%%%%%%%%%5
In a cold/warm partially ionized medium, a most relevant damping process is the ion-neutral damping \cite{Kulsrud-1971, Zweibel-1982-ion-neutral, Drury-1996-ion-neutral}.
For the typical conditions found in the disk of the MW, this process is very effective, so that the damping timescale for waves resonant with CRs of momentum $\lesssim 100$ GeV is of the order of tens of years. This means that, probably, the bulk of CRs stream very fast in the disk and is weakly coupled to the ISM. In the halo, instead, the fraction of neutrals is negligible and CRs can be scattered. The galactic halo is probably the actual diffusion region. In this scenario, the  transport of CRs in the disk region is unclear\cite{Skilling-1971, Holmes-1975}. On the other hand, another possibility is that the neutral phases are mostly confined in clumps/clouds, while most of the disk volume is filled with warm/hot ionized gas\cite{Ferriere-2001}, so that CR diffusion could work also in the near-disk region. 
This aspect is very relevant for the CR-driven wind formation, since the coupling between CRs and the ISM provided by the CR scattering on plasma waves, is necessary  to launch winds\cite{Breitschwerdt-1991-I, Recchia-winds-I}.\\

Starting from the CR and wave distribution functions one can define the corresponding hydrodynamic  energy density and pressure that determine the wind hydrodynamics:
\begin{align}
    &\epsilon_c(\vec{r}) = \int_0^\infty dp\; 4\pi p^2\, T(p)\,f(p, \vec{r}) ,\\
    &P_c(\vec{r}) = \int_0^\infty dp\; \frac{4\pi}{3}  p^2\, pv(p)\,f(p, \vec{r})\\ \label{eq:Pc-from-spectrum}
    &\epsilon_w(\vec{r}) = \int_{k_0}^\infty \frac{{B}^2}{4\pi}W\,(k, \vec{r})\,dk\\
    & P_w(\vec{r})=\epsilon_w/2,
\end{align}
where $T(p)=\sqrt{p^2c^2 + (mc^2)^2}-mc^2$ is the particle  kinetic energy.

\section{Cosmic ray driven winds: hydrodynamics}
\label{sec:wind-hydro}

The hydrodynamics of CR-driven winds is described by the equations for the conservation of mass, momentum and energy for the gas, by the evolution of the background magnetic field and  by the equations for the CR and magnetic wave pressure. The last are  derived by appropriate integrations, respectively  in momentum and in wave number, of  the transport equation of CR and of the waves (see Sec.~\ref{sec:wind-CR-transp}). 
The equations\cite{Breitschwerdt-1991-I} read
\begin{align}
&    \frac{\partial \rho}{\partial t} + \Vec{\nabla}\cdot \left(\rho \Vec{u} \right) = q\\
&   \frac{\partial \rho \Vec{u} }{\partial t} + \Vec{\nabla}\cdot \left( \rho\Vec{u}\Vec{u} + P_{tot}\mathbf{I} - \frac{\rho\Vec{B}\Vec{B}}{4\pi}   \right) = \rho\Vec{g} + \Vec{m}\\
& \frac{\partial e_{tot} }{\partial t} +
\Vec{\nabla}\cdot \Vec{S} = \rho\Vec{u}\cdot (\Vec{g} + \Vec{m}) + \epsilon \\
& \frac{\partial \Vec{B} }{\partial t} = \nabla \times \left(\Vec{u}\times \Vec{B}\right)\\
& \Vec{\nabla}\cdot \Vec{B} =0\\
& \frac{\partial}{\partial t} \left(\frac{P_c}{\gamma_c -1} \right) + \Vec{\nabla} \left[ \frac{\gamma_c}{\gamma_c-1}\left(\Vec{u} + \Vec{v_A} \right)P_c  - \frac{\Bar{D}}{\gamma_c-1}\Vec{\nabla}P_c \right] 
=  \left(\Vec{u} + \Vec{v_A} \right)\cdot \Vec{\nabla} P_c + Q_c \\
& \frac{\partial e_w}{\partial t} + \Vec{\nabla} \left[\left(\frac{3}{2}\Vec{u} + \Vec{v_A} \right)e_w \right] = \Vec{u}\cdot \Vec{\nabla} P_w - \Vec{v_A}\cdot \Vec{\nabla} P_c + Q_w, \label{eq:H-wave}
\end{align}
where 
\begin{align}
 & P_{tot} = P_g + P_c + P_w + \frac{B^2}{8\pi} \\
 & e_{tot} = \frac{1}{2}\rho u^2 + \frac{P_g}{\gamma_g - 1} + \frac{P_c}{\gamma_c - 1} + e_w + \frac{B^2}{4\pi} \\
 & \Vec{S} = \left(\frac{1}{2} u^2 + \frac{\gamma_g}{\gamma_g -1}\frac{P_g}{\rho}\right)\rho \Vec{u}  + \frac{1}{\gamma_c-1}\left[ \gamma_c\left(\Vec{u} + \Vec{v_A} \right)P_c - \Bar{D}\Vec{\nabla}P_c \right] \\ \nonumber
& \hspace{10pt} + \left(\frac{3}{2}\Vec{u} + \Vec{v_A} \right)e_w  + \frac{\Vec{E}\times \Vec{B}}{4\pi}.
\end{align}
Here $\rho$, $\Vec{u}$, $P_g$ and $\gamma_g$ are the gas density, velocity, pressure and adiabatic index, respectively. $P_c$, $\gamma_c$ and $\Bar{D}$ are the CR pressure, adiabatic index and momentum averaged diffusion coefficient. $\Vec{B}$ is the large scale background magnetic field and $e_w = 2P_w = \langle \delta \Vec{B}^2 \rangle /4\pi $ is the Alfv\'{e}n  wave energy density, whose speed is 
\begin{equation}
    \Vec{v}_A(\vec{r}) = \frac{\Vec{B}(\vec{r})}{\sqrt{4\pi\rho(\vec{r})}}.
\end{equation}
These equations may also include possible additional sources and sinks of mass, momentum and energy for the wind, deriving, for instance, from supernovae explosions and stellar winds, radiative  cooling and the damping of plasma waves. 
$g(\vec{r})$ represents the galactic gravitational acceleration.

The hydrodynamic CR and wave equations can be derived from the corresponding kinetic transport equations by integration  in momentum and wavenumber, respectively\cite{Recchia-winds-I}. 
The expression for the momentum  averaged diffusion coefficient is given by, 
\begin{equation}
    \bar{D}(\vec{r})=\frac{\int_0^\infty dp\; 4\pi  p^2\, T(p)D(p, \vec{r})\vec{\nabla}f}{\int_0^\infty dp\; 4\pi p^2\, T(p)\vec{\nabla}f},
\end{equation}
while the  CR adiabatic index is defined as $\epsilon_c = P_c/(\gamma_c -1)$ and $\gamma_c = 4/3 (5/3)$ for relativistic(non relativistic) particles.
It can be also  shown\cite{Recchia-winds-I} that the hydrodynamic counterpart of the streaming instability term in the wave transport equation, $\Gamma_{\rm CR}W$, is given by  $\vec{v}_A\cdot \vec{\nabla}P_c$, which appears in  the RHS of Eq.~\ref{eq:H-wave}.

In many cases relevant for CR propagation, the wave damping, also in the hot ionized medium of galactic halos,  is fast  enough that waves produced by streaming instability are practically damped locally and one can  safely ignore the actual wave transport\cite{Everett-2008}. In this case, the level of magnetic turbulence resonant with CR particles is determined by the balance between the growth and damping rate of waves, $\Gamma_{\rm CR} = \Gamma_{\rm damp}$. The resulting wave pressure is usually orders of magnitude smaller than that of CRs and of the background plasma and its contribution to the wind hydrodynamics can be neglected. However such turbulence is still necessary in order to scatter CRs and thus to couple them to the ISM. In addition, the dissipated waves heat the gas. This effect is taken into account by adding the  the term $\vec{v}_A\cdot \vec{\nabla}P_c$ to the  the equation for the gas energy density, since all the energy injected by CRs into waves is transferred to the gas. 

The hydrodynamic of CR-driven winds has been widely investigated in the literature, starting from the relatively computationally simpler and faster  stationary approaches\cite{Breitschwerdt-1991-I, Breitschwerdt-1993-II, Zirakashvili-1996-wind-rot-gal, Everett-2008, Recchia-winds-I, Recchia-winds-II}, to time-dependent calculations up complex, refined simulations\cite{Uhlig-2012, Booth-2013, Salem-2014, Girichidis-2016, Ruszkowski-2016, Pakmor-2016-wind-diffusion-simu, Pakmor-2016-wind-cr-transport-simu, Simpson-2016-cr-pressure-winds-simu, Wiener-2017-streaming-diffusion-simu, Pfrommer-2017-simu, Fujita-2018-wind-simu, Farber-2018-wind-simu, Jacob-2018-winds-halo-mass-simu, Thomas-2019-cr-hydro-simu}.

\section{Stationary cosmic ray-driven winds in the Milky Way}
\label{sec:wind-stat}

So far, the only self-consistent solutions of the coupled system of equations of the CR transport and of the CR-driven wind hydrodynamics, have been obtained solely in the stationary regime and under some simplifying assumptions, for instance on the large scale structure of the  galactic magnetic field, and neglecting radiative cooling\cite{Recchia-winds-I, Recchia-winds-II}.
Despite these limitations, the study of stationary winds provides precious information on their properties and on their impact on the CR transport. Moreover, being much easier and faster than simulations and of more complex time-dependent approaches,  it allows to quickly explore wide parameter ranges\cite{Everett-2008, Recchia-winds-II}. 

In this section we report a common  model of stationary winds, illustrating the main physical ingredients, the relevant launching parameters in the MW and some possible implications on the observed CR spectrum. 

\subsection{The geometry of the magnetic field lines}

Radio observations show the presence of galactic magnetic fields, usually  of the order of $\rm \mu G$ and  with a coherence length that can exceed $\sim$ kpc,  both in the MW and in other galaxies\cite{Pshirkov-2011-B-gal, Beck-2013-B-gal, Valle-2017-B-gal}. 
Their origin is commonly attributed to complex dynamo processes (see Refs.~\refcite{Widrow-2002-B-gal-origin}, \refcite{Hanasz-2017} and references therein). In the disk region, field lines are mainly parallel to the disk surface, while also X-shaped, out of disk components, have been detected\cite{Jansson-2012}.
The activity from several classes of astrophysical sources, such SNe explosions of single stars and cluster of stars, and  stellar winds, which also contribute to the formation of GWs, can push the field lines in the disk and make them break through the halo. The pressure support provided by these events may last $10^6-10^7$ yr, but the  overlap of the effect of multiple such events, randomly taking place in the galactic disk, may keep the field line opening up to distances from the disk of the order of the disk radius and for time scales of $\gtrsim 10^8-10^9$ yr.    
Moreover, in the presence of GWs, field lines attached to the outflowing material can be further transported outward.
The rotation of the galaxy additionally contribute to shape the large scale magnetic field\cite{Breitschwerdt-1991-I, Zirakashvili-1996-wind-rot-gal}.

In principle, the magnetic field geometry, which is also essential to CR propagation, should be determined self-consistently with the wind structure, which represents a big complication in models of CR-driven winds.
However a good approximation, adopted in several stationary approaches\cite{Breitschwerdt-1991-I, Everett-2008, Recchia-winds-I}, is to pre-assign the magnetic flux tube shape in a way that mimics the geometries found in more complex simulations, giving its area as a function of the distance from the disk. With such approach the equations also becomes one-dimensional and only depend on the coordinate perpendicular to the galactic disk. A typically adopted geometry, that intuitively encompass the ingredients illustrated above and confirmed by simulations,  is the mushroom-like shape in which the magnetic flux tube is roughly cylindrical up to a given distance from the disk (comparable with the disk radius) and then opens up in a roughly cylindrical way.
%%%%%%%%%%%%%%
\subsection{Stationary  equations in the given flux tube geometry}

In the stationary regime and assuming a given flux-tube geometry, as explained above,   the  equations greatly simplify and become one-dimensional\cite{Breitschwerdt-1991-I, Recchia-winds-I}. 
Be $z$ the coordinate perpendicular to the disk, and $A(z)$ the area of the mushroom-like flux tube
\begin{equation}
    A(z)=A_{0} \left[ 1+ \left( \frac{z}{Z_b}\right)^{\alpha}\right],
\end{equation}
where $Z_b \sim$ disk radius  and $\alpha = 2$ for  spherical opening. \\
The hydrodynamic equations and the CR transport equation become\cite{Breitschwerdt-1991-I, Recchia-winds-I}
\begin{align}      
& \rho u A=\textnormal{const}, \label{eq:H-mass} \\
& AB= \textnormal{const},\label{eq:H-B}\\
& \frac{du}{dz}=u\frac{c_{*}^2\frac{1}{A} \frac{dA}{dz} -\frac{d\Phi}{dz}}{u^2-c_{*}^2}, \label{eq:H-wind-eq}\\
& \frac{dP_g}{dz}=\gamma_g\frac{P_g}{\rho}\frac{d\rho}{dz}-(\gamma_g -1)\frac{v_A}{u}\frac{dP_c}{dz} \label{eq:H-Pg}\\
&\frac{dP_c}{dz}= \gamma_{eff}\frac{P_c}{\rho}\frac{2u+v_A}{2(u+v_A)}\frac{d\rho}{dz}, \label{eq:H-Pc}\\
& c_{*}^2=\gamma_g \frac{P_g}{\rho} + \gamma_{eff} \frac{P_c}{\rho} \left[ 1-(\gamma_g-1)\frac{v_A}{u} \right]\frac{2u+v_A}{2(u+v_A)}, \label{eq:H-cs}\\
& \frac{\gamma_{eff}}{\gamma_{eff} -1}  =  \frac{\gamma_{c}}{\gamma_{c} -1} -\frac{\overline{D}}{(\gamma_c -1)(u+v_A)P_c}\frac{dP_c}{dz} \label{eq:H-GAeff}\\
&\frac{u^2}{2} +\frac{\gamma_g}{\gamma_g -1} \frac{P_g}{\rho} + \Phi +\frac{\gamma_{eff}}{\gamma_{eff} -1}  \frac{P_c}{\rho}\frac{u+v_A}{u} = \textnormal{const} \label{eq:H-energy}
\end{align}
and
\begin{equation}\label{eq:kin-CR}
\frac{\partial}{\partial z} \left[ A\,D\frac{\partial f}{\partial z}\right] - A\,(u+v_A)\frac{\partial f}{\partial z}  +\frac{d\,A\,(u+v_A)}{dz}\frac{1}{3}\frac{\partial f}{\partial \ln p}+A\,Q=0,
\end{equation}
where $\Phi(z)$ is the galactic gravitational potential. It has been assumed that Alfv\'en waves produced by CR streaming instability are damped locally and heat the gas\cite{Everett-2008, Recchia-winds-I}. 
$Q(z,p)$ represents the CR  injection, which, assuming that the sources are mainly SNRs, can be approximated as a power-law  in momentum and to be concentrated in a thin disk $Q(z,p) \approx Q_0(p)\delta(z)$. Moreover, $Q_0 \propto \xi_{CR} \mathcal{{R}_{SN}}$, where $\xi_{CR}$ is the CR acceleration efficiency and $\mathcal{{R}_{SN}}$ is the SN explosion rate.

Eq.~\ref{eq:H-wind-eq} is the so-called \textit{wind equation}, which represents the acceleration of the wind fluid. It contains the generalized sound speed $c_{*}$, which includes the  sound speed, $\sqrt{\gamma_g P_g/\rho}$, modified by the CR pressure through a "CR sound speed" that depends on the Alfv\'{e}nic Mach number $u/v_A$. It the case in which Alfv\'en waves are not damped very quickly, one would have an equation for waves and a "wave  sound speed" in the definition of $c_{*}$\cite{Breitschwerdt-1991-I}.  $\gamma_{eff}$ is an effective adiabatic index that takes into account the CR diffusivity. It has been shown that in most cases $\gamma_{eff}\sim  1.1-1.2 $, while $\gamma_c \sim 4/3$\cite{Recchia-winds-II}.
The wave heating is encompassed by the term $-(\gamma_g -1)v_A/u\, dP_c/dz$ in Eq.~\ref{eq:H-Pg}. \\

The wind equation \ref{eq:H-wind-eq} presents different classes of solutions depending on whether  its numerator or its denominator, or both,  vanish\cite{Ipavich-1975, Breitschwerdt-1991-I} at a given  position $z$. 
The location where both vanish is called the \textit{sonic point} and it corresponds to the position where the flow velocity equals the compound sound speed, $u=c_{*}$. \\
The solution relevant for GWs is the one for which the flow is subsonic near the galactic disk, at the position $z_0$ where the boundary conditions are specified, then it is  accelerated outward  and smoothly transit, at the sonic point $z_c$,   to the supersonic regime. At large enough distance from the disk, all pressures tend to zero  and the wind reaches its terminal velocity, $u_{f}$. 
The boundary conditions to be given at $z_0$ are, for instance, the gas density and temperature, the CR pressure and the value of the background magnetic field, but not the  launching velocity $u_0$. In fact, the launching velocity is  determined by imposing the passage through the sonic point. Since all other quantities are fixed at $z_0$, this also fixes the mass and energy fluxes. 

For launching velocities smaller than $u_0$ the flow never becomes supersonic and there is a point where the numerator of the wind equation vanishes. These solutions are called \textit{breezes} and are characterized by a non vanishing gas pressure at infinity. For  launching velocities larger than $u_0$ one may either have unphysical solutions, where only the denominator vanishes at some point, or purely supersonic flows. \\

%%%%%%%%%%%%%%%%%%%%
\begin{figure}[h!]
	\centering
	\includegraphics[width=\columnwidth]{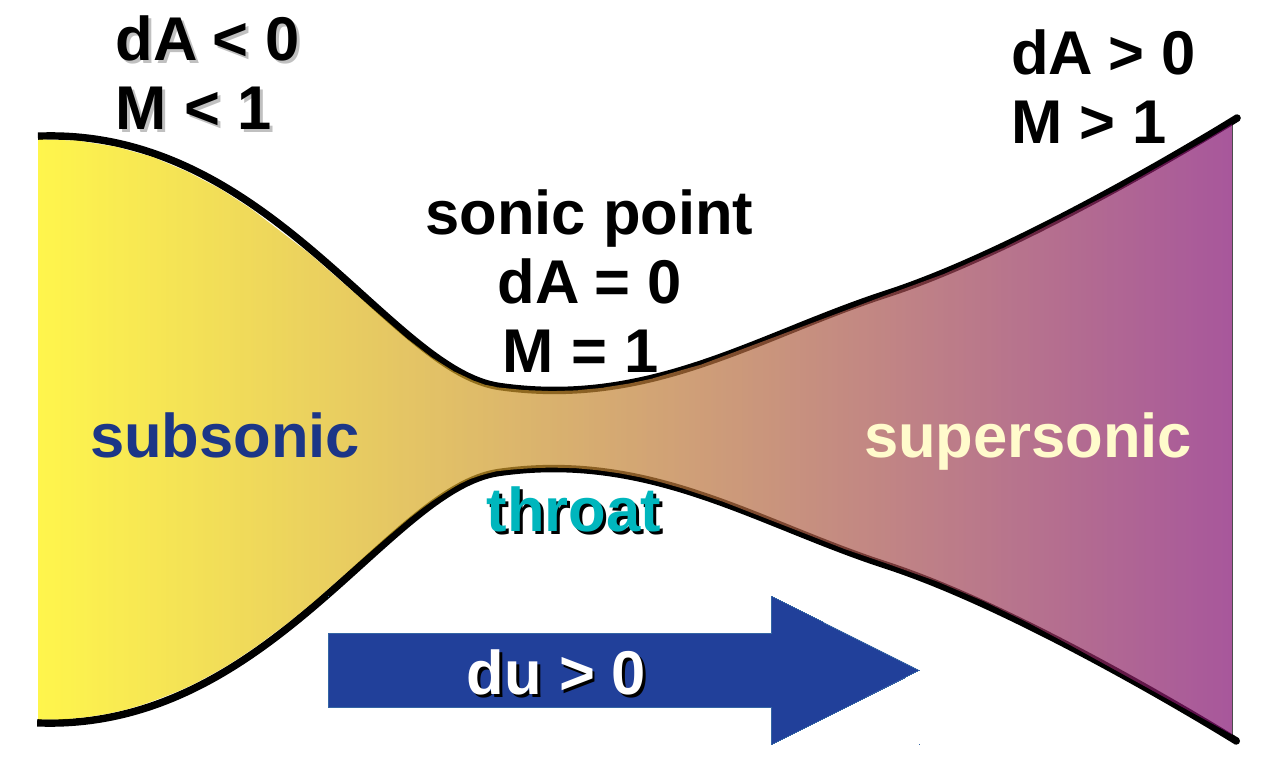}
    \caption{De Laval nozzle: the flow is constantly and smoothly accelerated all along the duct, from the subsonic regime (converging portion), to the supersonic regime (diverging portion). The sonic point coincide with the throat.}
	\label{fig:delaval}
\end{figure}
%%%%%%%%%%%%%%%%
The wind equation is analogous to that of the Parker model for the Solar wind\cite{Ipavich-1975, Parker-1965-solar-wind} and to that of the De Laval nozzle\cite{Recchia-winds-II}.
In fact, if one defines $a(z) \equiv \frac{1}{A}\frac{dA}{dz}$ and $g(z) \equiv \frac{d\Phi}{dz}$ in the wind equation, the last becomes formally identical to that of the De Laval nozzle
\begin{align}
& \frac{1}{u}\frac{du}{dz} = \frac{a(z) -\frac{g(z)}{c^{*2}}}{M^2 -1} \;\;\; (\textnormal{Wind equation}) \label{eq:dudz-wind} \\ 
\\ \nonumber
& \frac{1}{u}\frac{du}{dz} = \frac{\mathfrak{a}(z)}{M^2 -1} \;\;\; (\textnormal{De Laval nozzle equation})\label{eq:dudz-DeLaval}
\end{align}
where $M=u/c^{*}$ is the Mach number and $\mathfrak{a}(z) \equiv \frac{1}{A}\frac{dA}{dz}$ for a  nozzle of area $A(z)$, as shown in Fig.~\ref{fig:delaval}. The nozzles is composed by a duct that presents a converging section followed by a diverging section. The choke point corresponds to  $\mathfrak{a}(z) =0$. The geometry of the  nozzle allows a smooth transition from the subsonic to the supersonic regime. The flow begins as  subsonic in the converging portion, with $\mathfrak{a}(z) < 0$ and $M<1$, so that $du/dz >0$. At the throat $\mathfrak{a}(z) = 0$ and $M=1$ and the flow becomes transonic. Note that only in this way $du/dz$ can remain smooth. In the diverging portion $\mathfrak{a}(z) > 0$ and $M >1$ so that $du/dz$ remains positive and the flow continues to be accelerated (at least until a shock appears). \\
In the case of a wind the numerator $a(z) -\frac{g(z)}{c^{*2}}$ plays the role of the duct area in a De Laval nozzle. In order to keep accelerating the fluid away from the disk:
\begin{itemize}
	\item $u^2 < c^{*2}$ (subsonic flow)\\
	\\
	the gravitational term $g(z)$ has to  dominate. It behaves as a sort of "converging duct" term,\\ 
	\item $u^2 > c^{*2}$ (supersonic flow)\\
	\\
	the area term $a(z)$ has to  dominate. It behaves as a sort of "diverging duct" term.
\end{itemize}
This analysis illustrates that if the area of the flux tube is constant, $a(z)=0$, a stationary wind solution is not possible. Moreover, in an expanding flux tube, $a(z)>0$, as it happens in the case of a wind, the gravity allows for a stationary solution. 

\subsection{The role of cosmic rays and the cosmic ray spectrum}

The possibility to launch a (stationary) wind and its properties strongly depends on the conditions, such as the gas density and temperature, the CR density and the magnetic field  at the launching point near  the galactic disk. Moreover it also depends on the galactic gravitational potential (and in particular on the Dark Matter halo) and on the geometry of the field lines\cite{Recchia-winds-II}. 
As discussed above, the spatial dependent CR spectrum   affects the wind launching and is in turn affected by the presence of a wind. All these conditions depend on the type of galaxy and on the specific  location in  a galaxy. 

A parametric study of the wind launching is beyond the scope of this review, and the reader is referred to Refs.~\refcite{Everett-2008} and \refcite{Recchia-winds-II}, which performed a thorough investigation in the case of the MW. Moreover, Refs.~\refcite{Recchia-winds-I}  studied how the CR spectrum in the disk  gets modified when the launching conditions change and how this compares to the observed CR spectrum at Earth. Here we  briefly summarize the main findings of these studies, focusing specifically on the case of  MW and on winds launched  near disk region at the galactocentric distance of the Sun.    

\subsubsection{Wind hydrodynamics}

In general, stationary winds can exist only for closed intervals of the launching parameters. 
In the model presented here, the relevant quantities near the disk  are the ISM density, $n_0$, and temperature, $T_0$, the CR pressure, $P_{c0}$, and the magnetic field, $B_0$.\\
Winds probably originate from the less dense and hotter phase of the ISM, namely from the HIM\cite{Breitschwerdt-1991-I, Everett-2008}, which means for the MW, $n\approx 10^{-3}-10^{-2} \rm \, cm^{-3}$ and $T \approx \rm 10^6\, K$\cite{Ferriere-2001}. The CR pressure  is $P_c \sim 4\times 10^{-13}\, \rm erg/cm^3$\cite{Berezinskii-1990}. The galactic magnetic field is observed  to be of the order of few $\rm \mu G$\cite{Jansson-2012}. 
As for the flux tube area, $Z_b \approx 15$ kpc (the disk radius) and  $\alpha \approx 2.0$\cite{Breitschwerdt-1991-I, Recchia-winds-II}. \\
The gravitational potential of the MW is mainly contributed by the bulge, the disk and the DM halo. The last is especially relevant for GWs, since it extends for hundred kpc from the disk and affects the wind structure up to such large distances\cite{Recchia-winds-II}. \\
Wave damping is also a relevant ingredient in determining the possibility to launch winds\cite{Breitschwerdt-1991-I, Everett-2008, Recchia-winds-I}. 
In fact, in the absence of damping,  Alfv\'{e}n waves produced by CR streaming instability  can grow undisturbed, such  that their pressure becomes comparable to that of CRs and contribute to push the gas. This would make the wind launching easier, also in the case of denser and cooler gas. 
In the presence of wave damping the parameter space is additionally reduced by the fact that wave heating reduces the gas pressure gradient along $z$, and if it is too strong it may even stall the outflow\cite{Everett-2008}. 

\begin{figure}[h!]
	\centering
	\includegraphics[width=\columnwidth]{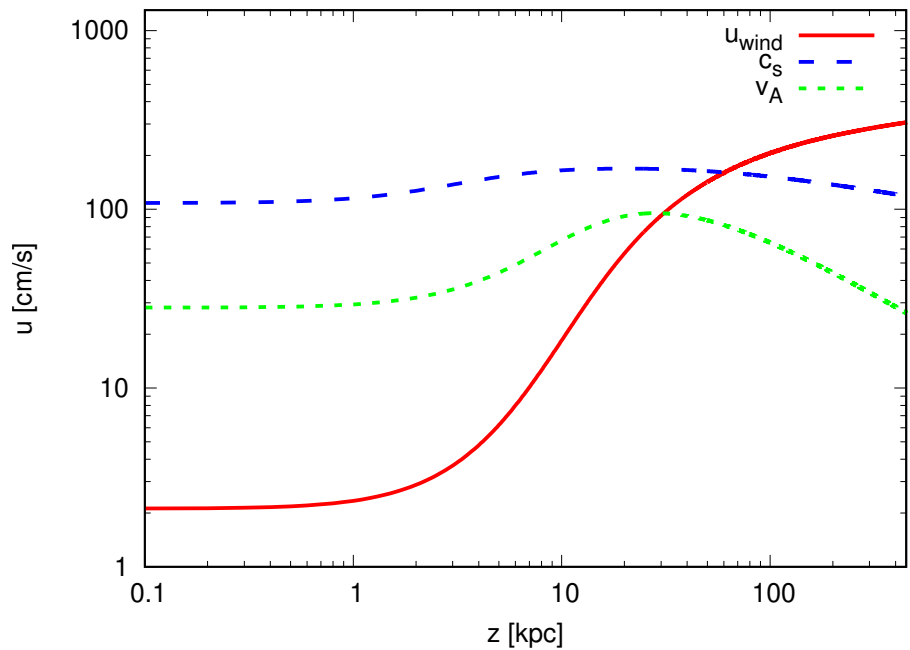}
    \caption{Example model: wind velocity $u_{wind}$, Alfv\'en velocity $v_A$ and compound sound speed $c^{*}$, as function of the distance from the disk. The sonic point is where $u_{wind}$ crosses $c^{*}$.}
	\label{fig:u}
\end{figure}

\begin{figure}[h!]
	\centering
	\includegraphics[width=\columnwidth]{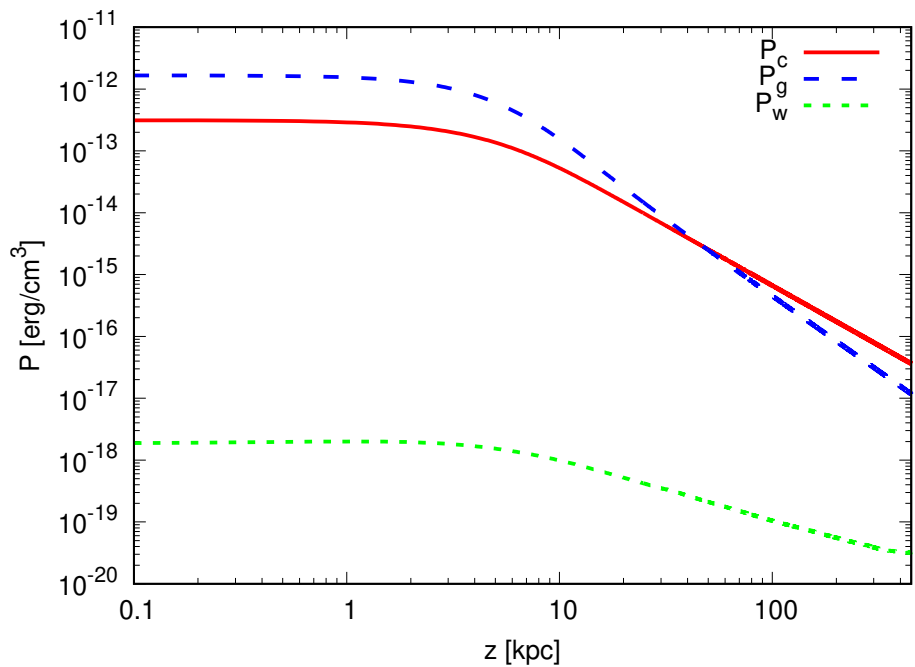}
    \caption{Example model: gas pressure $P_g$, CR pressure $P_c$ and wave pressure $P_w$, as function of the distance from the disk. Due to damping, the wave pressure is much smaller than the $P_g$ and $P_c$. Since the CR adiabatic index is smaller than the gas adiabatic index, $P_c$ drops with $z$ more slowly than $P_g$.}
	\label{fig:P}
\end{figure}

\begin{figure}[h!]
	\centering
	\includegraphics[width=\columnwidth]{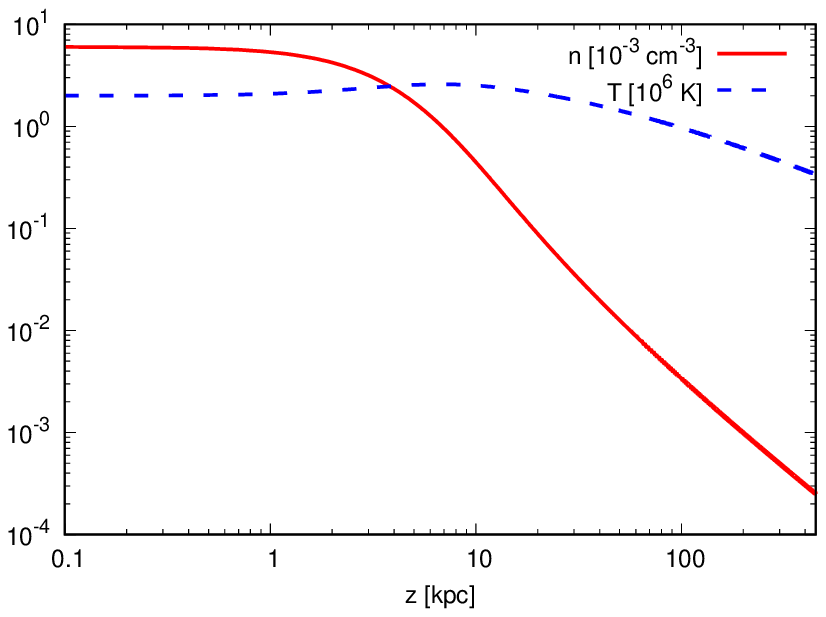}
    \caption{Example model: gas density $n$ and temperature $T$, as function of the distance from the disk. Due to wave heating, the gas temperature drops quite slowly with $z$. In fact the wind is roughly isothermal up to $\sim 100$ kpc from the disk.}
	\label{fig:n}
\end{figure}

It as been shown that, with launching parameters close to those mentioned above, stationary winds can exist. 
An example is reported in Fig.~\ref{fig:u}, \ref{fig:P} and \ref{fig:n}, where the chosen launching parameters are  listed in Table~\ref{tab:launch-param}.
\begin{table}[h!]
\tbl{Example of wind  launching parameters that are compatible with the observed conditions in the local Galactic environment and that allow to fit reasonably well the observed CR proton spectrum below $\sim 200$ GeV. The resulting  Alfv\'en and wind speeds at the launching point are $v_{A0} \sim 28$ km/s and $u_0 \sim 3$ km/s, respectively.  The resulting final velocity is $u_{f} \sim 404$ km/s, while the mass loss rate per unit area is $\Dot{m} \approx 4.7\times 10^{-4} \rm M_{\odot} kpc^{-2}yr^{-1}$.}
{\begin{tabular}{@{}ccc@{}} \toprule
Parameter & Symbol  & Value  \\ \colrule
gas density & $n_0$   & $6\times 10^{-3}$  \\
gas temperature & $T_0$  & $2\times 10^{6}$ K \\
magnetic field & $B_0$  & $1\, \mu G$ \\
magnetic flux tube scale height & $Z_b$  & 15 kpc\\
magnetic flux tube opening index & $\alpha $  & 2.0\\
CR pressure & $P_{c0}$  & $4\times 10^{-13}$\\ \botrule
\end{tabular} \label{tab:launch-param}}
\end{table}
The resulting  Alfv\'en speed is $v_{A0} \sim 28$ km/s, the wind launching velocity is $u_0 \sim 3$ km/s and the wind final velocity is $u_{f} \sim 404$ km/s, while the mass loss rate per unit area is $\Dot{m} \approx 4.7\times 10^{-4} \rm M_{\odot} kpc^{-2}yr^{-1}$. \\
This set of launching parameters  are compatible with the observed conditions in the local Galactic environment and lead  to a reasonably good fit of the observed CR proton spectrum below $\sim 200$ GeV, as illustrated in what follows.

Near the disk, the gas pressure dominates over the CR pressure and the flow is  sub-Alfv\'enic. A sizable fraction of the CR momentum is employed to produce Alfv\'en  waves, whose fast damping contribute to keep the gas hot up $\approx 100$ kpc and makes the gas temperature decrease with $z$ much slower than the gas density above $\sim 10$ kpc. 
On the other hand, CRs have a smaller adiabatic index than the gas ($\gamma_c = 4/3$, $\gamma_{eff} \approx 1.1-1.2$, $\gamma_g = 5/3$) so that the CR pressure decreases slower with $z$ than the gas pressure. 
This has prominent consequences  on the wind launching: the thermal pressure is effective at driving winds near the disk, but at larger $z$ CRs continue pushing  the gas, often making an outflow possible also in situations where the thermal pressure would not have been sufficient\cite{Everett-2008}.\\
%%%%%%%%%%%%%%%%%%%%
%%%%%%%%%%%%%%%%
%%%%%%%%%%%%%%%%%%%%
%%%%%%%%%%%%%%%%
%%%%%%%%%%%%%%%%%%%%
%%%%%%%%%%%%%%%%
It is interesting to notice that the mass loss rate per unit area reported above $\Dot{m} \approx 4.7\times 10^{-4} \rm M_{\odot}\, kpc^{-2}\,yr^{-1}$, if present on the whole Galactic  disk, would lead to a mass loss rate of $\Dot{M} \approx 0.36\, \rm M_{\odot}\, yr^{-1}$, which is comparable 
with the Galactic star formation rate ($\sim \rm M_{\odot}$/yr; see e.g Ref.~\refcite{Robitaille-2010}). This confirms the importance of CR-driven winds in MW-like galaxies.\\ 

For values of the launching parameters  much larger or smaller than the one quoted above,  winds might not exist or there could be some type of outflow but it would not be stationary. For instance, if the ISM is too dense and/or cool, the available energy could not be enough to drive an outflow, while for very hot and dilute gas the outflow may resemble more to an evaporation.\\
In general, an increase of the gas density leads to winds that have a  smaller launching velocity and a smaller  final velocity, and overall, a smaller mass loss rate.   An increase of the gas temperature and of the CR pressure lead to an increase of $u_0$, and, with fixed $n_0$, also to an increase of the mass loss rate. However, multiplying $T_0$ and $P_{c0}$ by the same factor, the resulting increase in $u_0$ is larger in the case of an increase  of $T_0$. 
This is due to the fact that momentum and energy deposited before the sonic point can affect both $u_0$ (and $\Dot{m}$) and $u_{f}$, while beyond the sonic point they affect only $u_{f}$, since $u_0$ is determined by the passage through the critical point. 
As illustrated above, the larger adiabatic index of the gas compared to CRs implies that the gas pressure get "exhausted" closer to the disk, while the CR pressure continues to be effective also at larger distances. Thus, the thermal pressure is generally more efficient at increasing the $u_0$ (and $\Dot{m}$).
Moreover, the effect of the CR pressure also depends  on whether the flow is sub-Alfv\'enic or super-Alfv\'enic. In fact, in these two limits the CR pressure gradient becomes\cite{Everett-2008, Recchia-winds-II}:
\begin{align}
\lim_{u << v_A} \frac{dP_c}{dz} = \frac{\gamma_{eff}}{2}\frac{P_c}{\rho} \frac{d\rho}{dz} \\
\nonumber \\
\lim_{u >> v_A} \frac{dP_c}{dz} = \gamma_{eff}\frac{P_c}{\rho} \frac{d\rho}{dz}.
\end{align}        
Thus, the effective CR adiabatic index is reduced in the sub-Alfv\'enic case, making CRs less efficient at driving the wind in this regime. 
The behavior of $u_f$ is more complex and in general an increase of $T_0$ and $P_{c0}$ does not necessarily lead to an increase of $u_f$. 
In fact, $u_f$ is computed from the energy conservation equation \ref{eq:H-energy}, considering that at large $z$  all other quantities drop to zero, 
and the total wind energy is solely kinetic:
\begin{equation}\label{eq:H-uf}
\frac{u_f^2}{2} =\frac{u_0^2}{2} +\frac{\gamma_g}{\gamma_g -1} \frac{P_{g0}}{\rho_0} + \Phi(z_0) +\frac{\gamma_{eff}}{\gamma_{eff} -1}  \frac{P_{c0}}{\rho_0}\frac{u_0+v_{A0}}{u_0} .
\end{equation}
The CR term depends on the Alfv\'{e}nic Mach number $M_{A0} = u_0/v_{A0}$ through the factor $(1+M_{A0})/M_{A0}$, which is large for  $M_{A0} \ll 1$ and becomes unity for $M_{A0} \gg 1$. 
An increase of either $T_0$ or $P_{c0}$ increases $u_0$ and, thus $M_{A0}$. Correspondingly, the $(1+M_{A0})/M_{A0}$  decrease. If such decrease is compensated by the increase of  $P_{g0}$ or $P_{c0}$, $u_f$ will increase, otherwise $u_f$ will decrease. In the limit in which the flow is super-Alfv\'enic $u_f$ increases with $T_0$ or $P_{c0}$\cite{Recchia-winds-II}. 

\subsubsection{Radiative cooling}

In the model illustrated here radiative cooling is not included. However it may be a key ingredient in shaping or hampering a wind and  deserves some attention. \\
For a hot ISM phase ($T\approx 10^6$ K), the dominant cooling process is the emission of forbidden lines and soft X-rays, at a rate given by\cite{Dalgarno-1972}
\begin{equation}\label{eq:cool-T}
\frac{1}{T}\frac{d\,T}{dt} \sim \Lambda(T)n^2,
\end{equation}
where the cooling function is $\Lambda(\rm T\, \sim\, 10^6) \sim 2\times \rm 10^{-23}erg\, cm^3/s$  for  Solar metallicity\cite{Dalgarno-1972}. 
The cooling rate increases with the gas density.\\
Radiative cooling can be  included in the stationary wind model, without affecting the topology of the possible solutions\cite{Breitschwerdt-1990-cool}  by modifying the equation of the gas pressure, Eq.~\ref{eq:H-Pg}, as
\begin{equation}\label{eq:cool-Pg}
\frac{dP_g}{dz}=\gamma_g\frac{P_g}{\rho}\frac{d\rho}{dz}-(\gamma_g -1)\frac{v_A}{u}\frac{dP_c}{dz} - (\gamma_g -1)\Lambda(T)\frac{n^2}{u},
\end{equation}
and the wind equation, Eq.~\ref{eq:H-wind-eq},  as
\begin{equation}\label{eq:cool-wind-eq}
\frac{1}{u}\frac{du}{dz}= \frac{c_{*}^2\, a(z) -g(z) + (\gamma_g -1)\Lambda(T)\frac{n}{u} }{u^2-c_{*}^2}.
\end{equation}
Comparing the timescale for  radiative   cooling  and  wave heating for the case shown in Fig.~\ref{fig:n}, one gets  $\tau_{cool} \sim 10^7$ yr and $\tau_{heat} \sim 10^8$ yr\cite{Recchia-winds-I, Recchia-winds-II}.
In general, in the absence of additional heating sources close to the disk, where the gas density is larger, radiative cooling might be so intense as to hamper the wind launching. A possible heating agent are  SN explosions which heat and ionize the surrounding medium and also inject magnetic turbulence that may eventually dissipate and additionally heat the gas\cite{Evoli-2018-halo}. Also ionization and  Coulomb losses of CRs may represent an important contribution to the gas heating\cite{Walker-2016}. 
Finally, radiative cooling may limit the "stationarity" of GWs, making the outflowing material to fall back to the disk\cite{Miller-2013, Miller-2015}, and establishing a gas recycling in the galaxy.   

\subsubsection{Cosmic ray spectrum}

When dealing with GWs launched at the galactocentric distance of the Sun, one has to keep in mind that, if on the one side the wind is determined by the CR pressure, the last depends on the CR spectrum,  which is affected by the wind and by self-generated waves in a highly non-linear way, and in the near-disk region should match the observed CR spectrum.\\

When we speak about the "observed CR spectrum", it is important to stress that the fluxes detected at Earth, for instance those reported by the AMS experiment\cite{amshard} are largely affected by the solar modulation at energies below $\sim 10$ GeV\cite{DiBernardo-2010}. On the other hand, the \textit{Voyager I} spacecraft has traveled far enough from the Sun that the measured spectrum, which is reported down to $\sim 3$ MeV, is no more affected by solar modulation\cite{Voyager-I-2013, Voyager-I-2016}.
On the other hand, the CR pressure is largely dominated by $\sim$ GeV CR protons, which represent the most abundant CR species (see Sec.~\ref{sec:CR-nutshell}). For this reason , the wind launching is mainly affected by  CRs whose local measurement is affected by the solar activity. Thus it is important to compare the computed  spectrum in the wind both with the \textit{Voyager I} data and with AMS data, provided that solar modulation is taken into account. \\

Sub-GeV CRs may still affect the wind through the heating produced by ionization and Coulomb losses\cite{Walker-2016}. 
The importance of low-energy CRs is not restricted to that. In fact, such CRs are widely recognized as the main ionization agent of the interior of  molecular clouds (see e.g Refs.~\refcite{Phan-2018-ionization}, \refcite{Recchia-2019-carrot} and references therein), with important implications on the interstellar chemistry and on star formation (see e.g \refcite{Krumholz-2011-ion-MC-star}). Moreover, they play a crucial role in the  nucleosynthesis of  light elements (lithium, beryllium, and boron)\cite{Tatischeff-Gabici-2018-review}.\\

The requirement that the calculated spectrum matches the observed spectrum provides a stringent constraint on the wind parameters\cite{Recchia-winds-I, Recchia-winds-II}. In fact, in many cases, winds launched with parameters compatible with the local ISM lead to CR spectra much different from the observed one. In most cases such spectra result harder than observations at low energy, below 100-1000 GeV, and much steeper than observations at higher energies\cite{Recchia-winds-I, Recchia-winds-II}.
In Refs.~\refcite{Recchia-winds-I} and \refcite{Recchia-winds-II} 
it is shown that, assuming that CR sources (say SNRs) inject a power-law spectrum in momentum, $Q_0(p) \propto p^{-\gamma}$, with $\gamma \approx 4.0-4.4$\cite{Blasi-2013-review, Aloisio-2015}, the overall normalization of the injection spectrum can be chosen in two ways, by suitably adjusting the SNe explosion rate (in the case of SNRs) and/or the CR acceleration efficiency: i) the calculated and observed proton spectra at Earth coincide at a specific particle momentum. Such momentum should be taken between 10-20 GeV since below that energy the spectrum detected at Earth is affected by solar modulation; ii) the total CR pressure computed from the proton  spectrum, using Eq.~\ref{eq:Pc-from-spectrum}, matches the observed one, $P_{c\,\odot} \sim 4 \times \rm 10^{-13}erg/cm^3$. Notice that, since the CR pressure derives from an integral in momentum  of the CR distribution, different CR spectra may lead to the same pressure. Moreover, since the CR spectrum is steeper than $\propto p^{-4}$, the specific spectral shape above $\sim 10$ GeV barely affects the total CR pressure.\\

\begin{figure}[h!]
	\centering
	\includegraphics[width=\columnwidth]{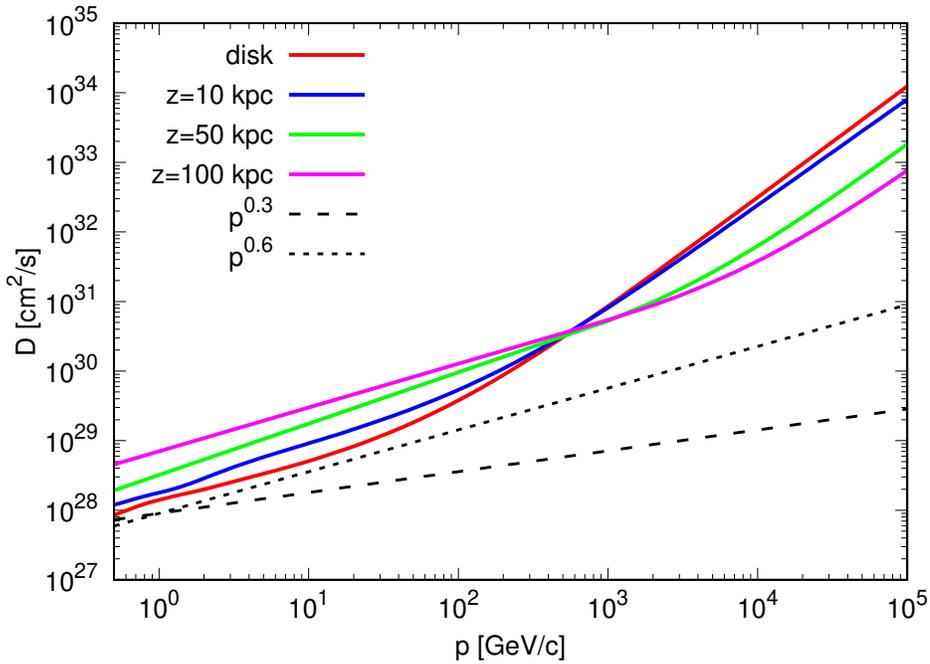}
    \caption{CR diffusion coefficient as function of the particle momentum $p$, for different distances from the disk. The corresponding wind structure is shown in Fig.~\protect\ref{fig:u}. Notice the sharp momentum dependence of $D(p)$, as compared to the $\propto p^{0.3-0.6}$ usually assumed in standard CR propagation scenarios\protect\cite{Berezinskii-1990, Blasi-2013-review}. This is a normal consequence of streaming instability, for which the level of waves depends on the CR density.}
	\label{fig:Dp}
\end{figure}

With both prescriptions on the normalization of the CR injection spectrum there is no guarantee at all that the computed spectrum matches the observed spectrum at all energies\cite{Recchia-winds-II}. 
This is mainly due to two factors which can be better understood keeping in mind the benchmark propagation model illustrated in Sec.~\ref{sec:CR-nutshell} and again assuming a power-law CR injection spectrum as illustrated above. For particle momenta for which advection dominates the spectral slope of the calculated  spectrum will be close to the injection slope, while, where diffusion dominates,  the spectral slope will be steeper. 
In the model presented in this section, the CR diffusion coefficient is solely due to self-generated Alfv\'en waves, whose growth rate, that depends on the CR density gradient,  decreases with energy (the CR spectrum is  steeper that $\propto p^{-4}$). This lead to a diffusion coefficient which sharply rises with energy, with a slope larger than the $\sim 0.3-0.6$ typically assumed in standard models of CR propagation\cite{Berezinskii-1990}, as shown in Fig.~\ref{fig:Dp}. 
%%%%%%%%%%%%%%%%%%%%
\begin{figure}[h!]
	\centering
	\includegraphics[width=\columnwidth]{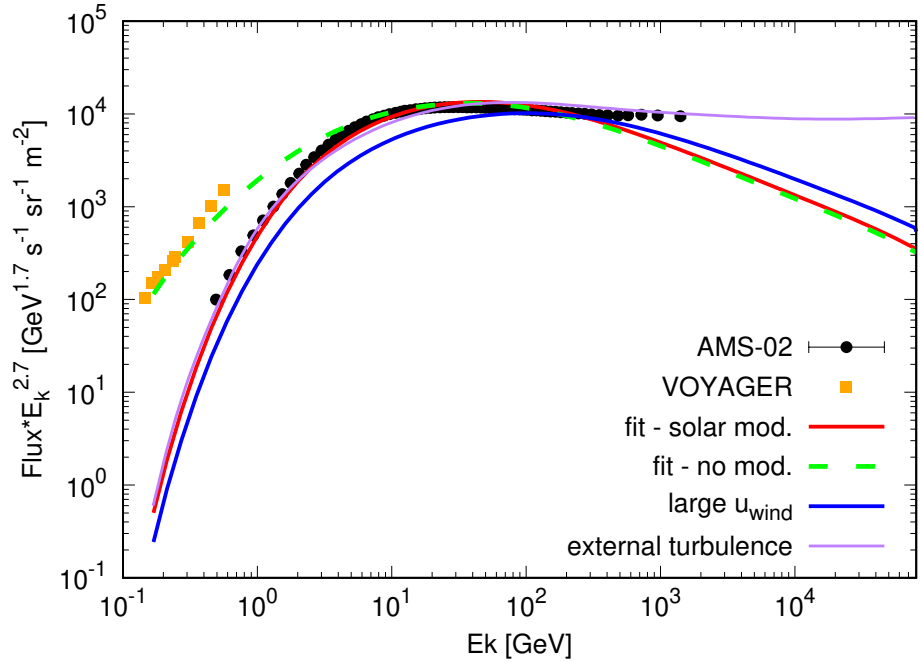}
    \caption{CR proton spectrum as a function of the particle kinetic energy in the near disk region, as compared to observations. The green curve shows the spectrum corresponding to the wind model of Fig.~\protect\ref{fig:u}, with a total  advection velocity near the disk of $\sim 30$ km/s. The red curve shows the same spectrum with the application of solar modulation\protect\cite{DiBernardo-2010}. The blue curve shows a spectrum obtained with a total advection velocity near the disk of $\sim 100$ km/s, which is too hard to fit data. The magenta curve shows qualitatively the effect of an additional, Kolmogorov-like, extrinsic turbulence that scatter CRs. This leads to a much harder spectrum at energies $\gtrsim 100-1000 $ GeV, as compared to streaming instability.}
	\label{fig:spectra}
\end{figure}
%%%%%%%%%%%%%%%%
%%%%%%%%%%%%%%%%%%%%
\begin{figure}[h!]
	\centering
	\includegraphics[width=\columnwidth]{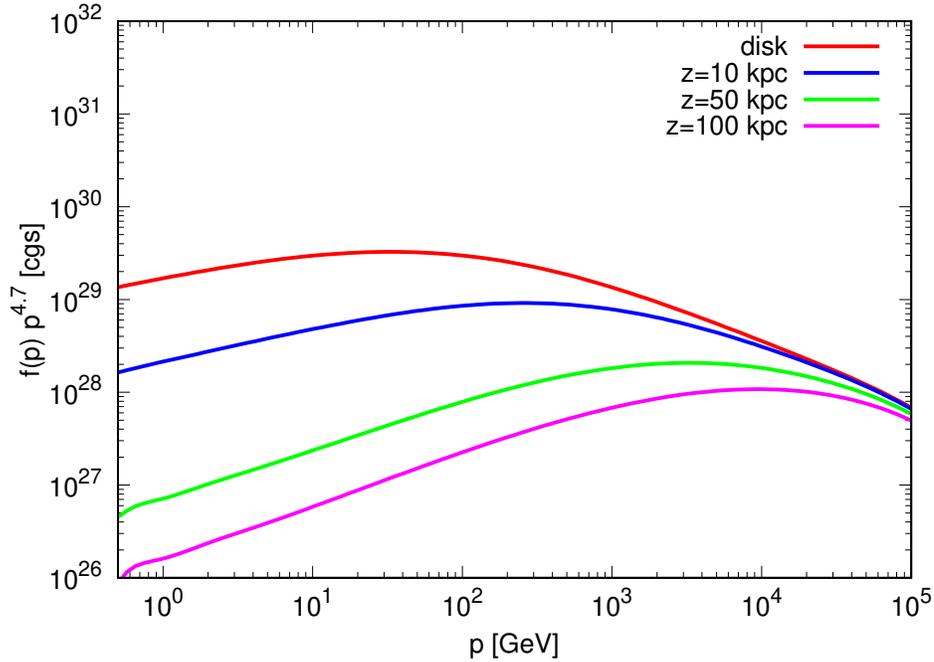}
    \caption{CR distribution as function of the particle momentum $p$, for different distance from the disk. The corresponding wind structure is shown in Fig.~\protect\ref{fig:u}.}
	\label{fig:fp}
\end{figure}
%%%%%%%%%%%%%%%%
%%%%%%%%%%%%%%%%%%%%
%%%%%%%%%%%%%%%%
%%%%%%%%%%%%%%%%%%%%
\begin{figure}[h!]
	\centering
	\includegraphics[width=\columnwidth]{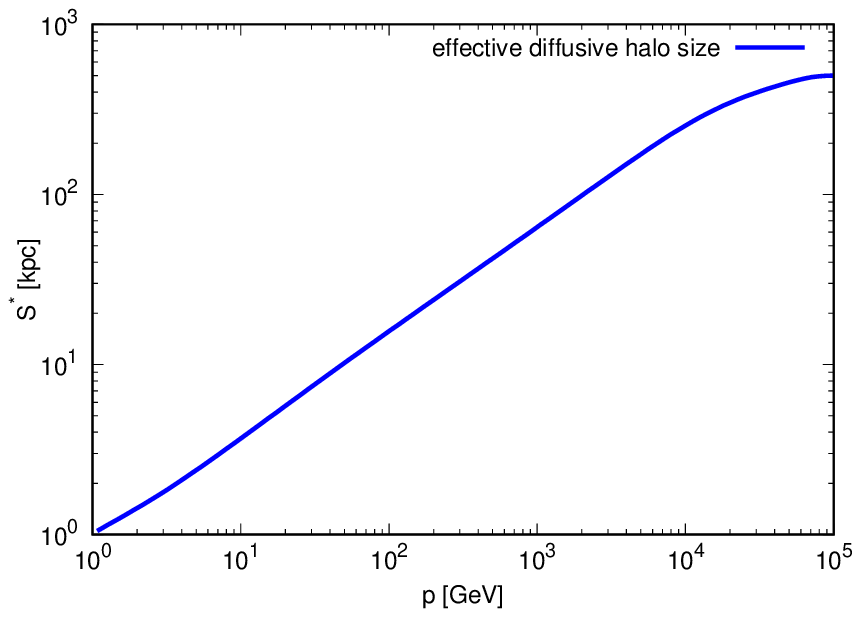}
    \caption{Effective diffusive halo size as function of the particle momentum $p$. The corresponding wind structure is shown in Fig.~\protect\ref{fig:u}. Particles of momentum $p$ are basically advected out of the galaxy for $z > S^{*}(p)$. The effective diffusive halo size increases with energy, as shown in Sec.~\protect\ref{sec:CR-nutshell}}
	\label{fig:Sp}
\end{figure}
%%%%%%%%%%%%%%%%
As illustrated above, winds are mainly launched from relatively hot and dilute phases of the ISM. This leads to a CR total advection velocity, $u +v_A$, close to the disk which can easily exceed $\sim 100$ km/s. Such value increases with $z$. The resulting CR transport will be dominated by advection up to hundreds GeV, resulting in a spectrum harder than observations at these energies.
A smaller advection speed can be obtained, at fixed $T_0$, by considering launching parameters still compatible with observations but with larger gas density and smaller $B_0$. 
This can be seen in Fig.~\ref{fig:spectra}, where we show the spectrum obtained with the launching parameters: $n_{0}=0.003$ cm$^{-3}$ and $B_{0}=2\,\mu $G, which gives $u_0 + v_{A0} \sim 100$ km/s; $n_{0}=0.006$ cm$^{-3}$ and $B_{0}=1\,\mu $G, which gives $u_0 + v_{A0} \sim 30$ km/s (the last values have been used in Fig.~\ref{fig:u}, \ref{fig:P} and \ref{fig:n}, and are reported in  Table~\ref{tab:launch-param}). In the case of larger density and smaller $B_0$, the predicted spectrum matches reasonably well observations, while with the other set of parameters the predicted spectrum is too hard up to hundreds GeV. This is a rather general feature of CR-driven winds, in which launching parameters compatible with observations and for which a wind can exist, may lead to a CR spectrum qualitatively different from the observed one, even if the corresponding CR pressure is close to observations.\\
For this reasons, the measurement of the CR spectrum at energies below few hundreds GeV represents a precious constraint on the wind parameters. This is surely true for winds launched at the galactocentric distance of the Sun, where we have a direct measurement of the CR spectrum, but it is applicable also to the cases in which an indirect indication of the CR spectrum is available, for instance thanks to $\gamma-$ray observations.
This is the case for CRs interacting with the ISM in the Galactic disk at different galactocentric distances\cite{Recchia-2016-grad}, or for $\gamma-$ray observations in other galaxies\cite{Peretti-2019-starbust, Krumholz-2020-starburst}. 

At higher CR energies ($E\gtrsim 100-1000$ GeV), the steep momentum dependence of the CR-induced diffusion coefficient, additionally enhanced by the opening of the flux tube, and only mildly compensated by the progressive increase of the effective diffusion region with energy (see Sec.~\ref{sec:CR-nutshell} and Fig.~\ref{fig:Sp}), gives a very steep spectrum. 
This result cannot be changed by a change of the launching parameters.
On the other hand, it has been shown by several authors (see e.g Ref.~\refcite{Aloisio-2015} and \refcite{Evoli-2018-halo}), that additional sources of turbulence, other than self-generated Alfv\'en waves, maybe injected by SNe explosions and characterized by a Kolmogorov-like spectrum (which would correspond to a diffusion coefficient $D\propto p^{0.3}$), may be relevant at those energies and could even revert the spectral trend illustrated here and lead to a hardening\cite{Aloisio-2015}. Such hardening is in fact measured in the CR spectrum (see Sec.~\ref{sec:CR-nutshell}). The possible effect of such an additional turbulence is qualitatively  shown in Fig.~\ref{fig:spectra}.\\
In Fig.~\ref{fig:fp} we show the CR distribution as a function of particle momentum $p$ for different distances from the disk. Low energy particles, below $\sim 100$ GeV, experience a diffusive halo smaller than $\sim 10$ kpc. Above that distance they are advected out of the galaxy in a progressively larger flux tube. This results in a fast decrease of the CR density  with $z$ and in a  progressively harder spectrum. Higher energy particles experience a much larger diffusive halo (see Fig.\ref{fig:Sp}). This leads to a smaller decrease of the CR density with  $z$. For such energies, the opening of the flux tube is partially compensated by a decrease of the diffusion coefficient at large $z$, also shown in Fig.~\ref{fig:Dp}. This effect is the result of the non-linear nature of the streaming instability combined with the decrease of the background magnetic field. If fact, if on the one side the CR density is small at high energies and slightly decreases with $z$, the background magnetic field also decreases, so that $\rm \langle \delta B \rangle /B$ in Eq.~\ref{eq:D-QLT} may become larger. This may in fact lead to an overall decrease of the diffusion coefficient, despite the small amount of produced waves\cite{Recchia-2016-grad}.

\section{Conclusions and perspective}
\label{sec:concl}

GWs represents a most important feedback process in the life of galaxies,  and their study constitutes  an active research area in astrophysics, both from the observational and theoretical point of view. 
GWs may be powered by several mechanisms, depending on the specific environment  in which they get formed. 

The key role of CRs in launching GWs, especially in MW-like galaxies, has been confirmed by many hydrodynamic studies.  
The coupling between CRs and the ISM, necessary for CRs to effectively push on the background plasma, is mediated by the scattering of CRs off plasma waves at the very small scales (well below 1 pc, compared to the multi kpc scale of the wind structure)  of the  Larmor radius of the CR particles in the galactic magnetic field.  Such scattering is also responsible for the diffusive motion of CRs.
The dynamics of CR-driven winds is a multiscale  non-linear phenomenon  intrinsically connected with the microphysics of the CR transport. In fact, the CR distribution responsible for the wind launching  is often affected by CR-induced plasma instabilities, by the (non-linear) damping of plasma waves,  and by the presence of the wind itself.  Moreover, the details of the CR propagation in different galactic environments  is also not well understood.
For these reasons,  CR-driven winds are particularly difficult to study, and most of current models only focuses on the wind hydrodynamics, treating CR as a relativistic gas and neglecting the kinematics of the CR propagation.  So far, the only self-consistent modeling of the CR propagation together with the wind structure has been obtained solely for CR protons, in the stationary regime, and under the assumption that only self-generated waves contribute to diffusion and with a pre-assigned geometry of the large-scale magnetic field. Moreover possible relevant effects,  such as radiative cooling, the effect of extrinsic sources of turbulence, and  ionization and Coulomb losses for sub GeV CRs, have been neglected.\\
Despite these limitations,  all these studies  allowed to investigate the wind structure in a variety of galactic environments and to  drive some general conclusions on the CR spectrum expected in the presence of GWs. In particular it has been shown that the measurement of the CR spectrum below few hundreds GeV represents an important constraint on the wind structure, since winds launched with parameters compatible with a given galactic environment generally lead to spectra different from observations.\\

Future improvements of the state of the art should focus on several aspects.\\
First of all, it is of primary importance a better understanding of the sources and types of plasma  turbulence that may scatter CRs, in addition to self-generated Alfv\'en waves, together with a deeper  study of the possible damping processes. For instance, the plasma turbulence injected by sources in the disk and cascading to smaller scales may represent a relevant contribution to CR scattering\cite{Evoli-2018-halo}, whose transport in the halo may be enhanced by GWs. This would require to solve also the wave transport equation.  The relevance of such investigation is obviously not restricted to CR-driven winds, and is central to the physics of CRs in general. \\
Ionization and Coulomb losses are important in determining the shape of the CR spectrum below GeV and may represent and important source of heating in the near-disk region\cite{Walker-2016}, which may mitigate the possible effects of radiative cooling\cite{Breitschwerdt-1991-I, Recchia-winds-I}.\\
Another important aspect is the transport of heavier nuclei in the wind, with possible relevant implications for the production of secondaries, and thus on observables
such as the B/C ratio. This would require the inclusion of spallation and nuclear decay terms in the equations.\\
An other interesting possibility could be the  reacceleration of CRs at the wind termination shock, analogous to the CR acceleration at SNR shocks\cite{Zirakashvili-2006-term-shock, Merten-2018-term-shock}.\\

Second, of great importance is to treat the full, hydrodynamic and CR transport problem, in time-dependent calculations. In fact, while in some interesting cases the assumption of stationarity  may be justified, it is inadequate in describing interesting outflows such as "galactic fountains", where the ejected material cools down and falls back to the disk, or outbursts produced in active galaxies, possibly with the contribution of CRs. 
Non stationary winds may originate from the Galactic center region, due e.g to a periodic activity of the central supermassive  black hole, and possibly connected with the \textit{Fermi Bubbles}\cite{Cheng--Fermi-Bubbles}.
Moreover, in time-dependent winds, the formation of multiple shocks may additionally contribute to the CR reacceleration.
Of great interest is also the role of CRs  in shaping the large scale, rotating, galactic magnetic field, which in turn affects the wind structure and the CR propagation\cite{Hanasz-2017}.\\
In general, the development of a fully  time-dependent, self consistent, CR-driven wind model, would allow to investigate such winds and the resulting CR distribution in detail and in a large variety of environments, from MW-like galaxies, to starburst\cite{Peretti-2019-starbust, Krumholz-2020-starburst} and AGN galaxies.\\

Third, GWs populate the galactic halos with hot plasma up to hundreds of kpc from the Galactic disk\cite{Nicastro-2012-gas-halo, Miller-2013, Miller-2015}, which act as a target for the interaction of CRs, with the consequent production of $\gamma-$rays and neutrinos. This may lead to a relevant contribution to the isotropic $\gamma-$ray background\cite{Feldmann-2013-gamma-back} and to the diffuse neutrino flux detected by Ice-Cube\cite{Taylor-2014-neutrini}. Such and extended  $\gamma-$ray halo has indeed been detected in the Andromeda galaxy\cite{Moskalenko-2019-M31-halo}. \\

As a final remark, it is important to stress the possible prominent role played by CRs in the evolution of galaxies. 
In this review we showed the effect of CR driving  in MW-like galaxies, where the mass loss  induced by CR-driven winds can be comparable to the mass formed in stars. Despite the role of CRs in starburst and AGN galaxies is not that clear, it may also be relevant. This, together with the other possible effects of CRs on the galactic environments discussed above, shows the importance of including CRs in models of galactic evolution.

%%%%%%%%%%%%%%%%%%%%%%%%%%%%%%%%%%%%%%%%%%%%%%%%%%%%%%%%%%%%%%%%%%%%%
%%%%%%%%%%%%%%%%%%%%%%%%%%%%%%%%%%%%%%%%%%%%%%%%%%%%%%%%%%%%%%%%%%%%%
%%%%%%%%%%%%%%%%%%%%%%%%%%%%%%%%%%%%%%%%%%%%%%%%%%%%%%%%%%%%%%%%%%%%%

\section*{Acknowledgments}

The author acknowledge support from Agence Nationale de
la Recherche (grant ANR- 17-CE31-0014). 

\bibliographystyle{ws-ijmpd}
\bibliography{biblio}

\end{document}